\documentclass[12pt]{article}
\catcode`\@=11
\@addtoreset{equation}{section}

\global\arraycolsep=2pt
\usepackage{mathrsfs,amsbsy,amssymb,latexsym,amsfonts,amsmath,cite}
\usepackage{graphicx,color}
\usepackage{physics}
\usepackage{mathtools}
\usepackage{ulem}
\usepackage{caption}
\usepackage{here}

\usepackage{mathrsfs,amsbsy,amssymb,latexsym,amsfonts,amsmath,cite,bm}
\usepackage{graphicx,color}
\usepackage{mathtools}
\usepackage{enumerate}

\DeclareFontFamily{U}{BOONDOX-calo}{\skewchar\font=45 }
\DeclareFontShape{U}{BOONDOX-calo}{m}{n}{
  <-> s*[1.05] BOONDOX-r-calo}{}
\DeclareFontShape{U}{BOONDOX-calo}{b}{n}{
  <-> s*[1.05] BOONDOX-b-calo}{}
\DeclareMathAlphabet{\mathcalboondox}{U}{BOONDOX-calo}{m}{n}
\SetMathAlphabet{\mathcalboondox}{bold}{U}{BOONDOX-calo}{b}{n}
\DeclareMathAlphabet{\mathbcalboondox}{U}{BOONDOX-calo}{b}{n}

\newcommand{\cC}{\mathcal C}

\allowdisplaybreaks

\begin{document}

\renewcommand{\thefootnote}{\fnsymbol{footnote}}

\begin{flushright}
KUNS-2922
\end{flushright}
\vspace*{0.5cm}

\begin{center}
{\Large \bf  Chaotic instability in the BFSS matrix model
}
\vspace*{2cm} \\
{\large  Osamu Fukushima$^{\sharp}$\footnote{E-mail:~osamu.f@gauge.scphys.kyoto-u.ac.jp}
and Kentaroh Yoshida$^{\sharp}$\footnote{E-mail:~kyoshida@gauge.scphys.kyoto-u.ac.jp}} 
\end{center}

\vspace*{0.4cm}

\begin{center}
$^{\sharp}${\it Department of Physics, Kyoto University, Kyoto 606-8502, Japan.}
\end{center}

\vspace{2cm}

\begin{abstract}
Chaotic scattering is a manifestation of transient chaos realized by the scattering with non-integrable potential. When the initial position is taken in the potential, a particle initially exhibits chaotic motion, but escapes outside after a certain period of time. The time to stay inside the potential can be seen as lifetime and this escape process may be regarded as a kind of instability. The process of this type exists in the Banks-Fischler-Shenker-Susskind (BFSS) matrix model in which the potential has flat directions. We discuss this chaotic instability by reducing the system with an ansatz to a simple dynamical system and present the associated fractal structure. We also show the singular behavior of the time delay function and compute the fractal dimension. This chaotic instability is the basic mechanism by which membranes are unstable, which is also common to supermembranes at quantum level. 
\end{abstract}

\setcounter{footnote}{0}
\setcounter{page}{0}
\thispagestyle{empty}

\newpage

\tableofcontents

\renewcommand\thefootnote{\arabic{footnote}}

\section{Introduction}

Chaos is a significant characteristic of nonlinear dynamics. 
Classically chaotic systems exhibit highly complex trajectories and the dynamics cannot be solved analytically. The complexity can be measured by  positive Lyapunov exponents and non-quasiperiodic plots in Poincar\'e section.
Recently, there has been an attempt to figure out chaotic behavior in the scattering process  \cite{Rosenhaus:2020tmv,Gross:2021gsj,Rosenhaus:2021xhm}. 
The S-matrix, which is one of the most important physical quantities, is a good target to tackle chaotic behavior in the context of field theory. In 
fact, chaotic behavior in the scattering process of non-integrable dynamical systems has long been well studied. This is called ``{\it chaotic scattering}''. 

\medskip 

What is the difference between the usual classical chaos and chaotic scattering? Classical chaos is usually characterized by long-time behavior. Therefore, we need to confine the system or consider an energy region where all motions are restricted to a finite region. However, there is another kind of chaos
which appears for finite time. This is called transient chaos. It may sound like a recurrence, but chaotic behavior can be clearly distinguished, for example, by making sure that the spectrum is continuous. Chaotic scattering is a manifestation of this transient chaos. 

\medskip 

How can one see this transient chaos? 
Let us consider a non-integrable system with a finite scattering region. Then a finite-time chaotic behavior is observed during a particle is passing through (or staying in) the scattering region, while the particle is free outside the scattering region. This is nothing but the transient chaos that is identified as the chaotic scattering. This chaotic scattering has been investigated in various models such as the H\'enon-Heiles system and the Contopoulos system. It also shows stretching and folding of the phase space and consequently leads to a horse-shoe construction of stable manifolds. This indicates that the chaotic scattering should also have a fractal structure.

\medskip

It is well known that classical Yang-Mills theory exhibits classical chaos \cite{Matinyan:1981dj}. Similarly, the Banks-Fischler-Shenker-Susskind (BFSS) matrix model \cite{Banks:1996vh}, which is a one-dimensiomal matrix quantum mechanics, is also chaotic \cite{Arefeva:1997oyf}\footnote{The Berenstein-Maldacena-Nastase (BMN) matrix model \cite{Berenstein:2002jq}, which is a kind of massive deformation of the BFSS matrix model, also shows classical chaos \cite{Asano:2015eha}. In this case, the flat directions in the BFSS matrix model are lifted up by mass terms and thus all of the classical motions are completely bounded.}. The BFSS matrix model is a matrix realization of light-front M-theory and a matrix regularization of the supermembrane theory \cite{dWHN}. The potential of the BFSS matrix model has flat directions and a motion along this direction may escape from the chaotic region to a spatial infinity. Hence the BFSS matrix model may exhibit transient chaos. These flat directions are considered as the origin by which  supermembranes are unstable at quantum level \cite{dWLN}. 

\medskip

In this paper, we discuss chaotic scattering in the context of the BFSS matrix model. By reducing the model to a simple dynamical system with a specific ansatz, we study the fuzzy membrane dynamics. The initial configuration initially moves in the chaotic region of the potential, but extends infinitely long along a flat direction after a certain period of time. This infinitely long extension, which indicates membrane instability \cite{dWLN}, corresponds to the escape process in chaotic scattering. Once this instability is recognized as a chaotic scattering process, we can present the associated fractal structure 
and compute the life time of the configuration.  
The singularities in the time delay function form a Cantor-like set and its fractal dimension can be estimated by means of uncertain fraction. 
Thus a fractal structure behind the BFSS matrix model is unveiled by reformulating the membrane instability \cite{dWLN} as a chaotic scattering process. 

\medskip

This paper is organized as follows. Section \ref{sec:transient} provides an overview of transient chaos.
In Section \ref{sec:four-hill} we consider a system with a four-hill potential as a concrete example. A fractal structure appears in the space of initial conditions for which trajectories survive at a certain time $t$\,. The singularities in the time delay function forms a Cantor-like set with a fractal dimension. In Section \ref{sec:matrix}, by reducing the BFSS matrix model to a simple dynamical system, the membrane instability is reinterpreted as the chaotic scattering. The analysis is almost parallel to the four-hill case. Section \ref{sec:conclusion} is devoted to conclusion and discussion.

\section{An overview of transient chaos}\label{sec:transient}

Let us here give an overview of transient chaos in a general setup.
For good reviews, for example, see  \cite{scattering-review,transient book}. Some of the results in \cite{scattering-review} are reproduced by our own program. 

\subsection*{General characteristics of classical chaos}

Classical chaos is usually discussed in a bounded system 
in order to study long-time behaviors of trajectories. 
A chaotic dynamical system has the following two characteristics:
\begin{enumerate}
\item strong sensitivity to initial conditions,
\item fractal from a symbolic dynamics.
\end{enumerate}
The first property can be measured by the Lyapunov spectrum.
Suppose that a solution $\bm{x}(t)$ is chaotic and consider a nearby point $\bm{x}(t) + \delta\bm{x}(t)$\,.
Then the deviation $\delta\bm{x}(t)$ grows exponentially like
\begin{align}
|\delta\bm{x}(t)|\sim |\delta\bm{x}(0)|\,e^{\lambda t}\,.
\end{align}
The constant parameter $\lambda$ is a Lyapunov exponent. This means that tiny differences in the initial conditions will grow exponentially over time. This property is often called the butterfly effect. 

\medskip

The second means that a fractal structure exists behind classical chaos and this fractal arises from a symbolic dynamics generated by chaotic trajectories. Then the trajectories give rise to a Smale horseshoe.
The horseshoe construction leads to fractal structures.

\subsection*{Transient chaos}

{\it Transient chaos} is a phenomenon where a system exhibits chaos over {\it finite} time, while traditional chaos is rather characterized by a {\it long}-time behavior. When a non-integrable system is defined in an asymptotically flat space and has a finite scattering region, a finite-time chaotic behavior is observed during a particle is passing through (or staying in) the scattering region, while the particle is free outside the scattering region. This is an example of transient chaos that is identified as the {\it chaotic scattering}. 

\medskip

In comparison to the usual chaos, it is not so suitable to study the Lyapunov exponent because a particle will escape from the scattering region. 
However, this escape process is a characteristics of chaotic scattering and one can consider other quantities. 
For example, when one sees the space of initial conditions for which 
trajectories remain in the scattering region, there exist some gaps which corresponds to escaped trajectories. Then the gaps define topology. The number of gaps increases exponentially over time like 
\begin{align}
N(t)\sim N(0)\,e^{ht}\,. 
\label{topological entropy}
\end{align}
Here $h$ is referred to as the topological entropy because the escaped trajectories imply the information loss.  

\medskip 

In addition, the gaps show a self-similar structure, which is nothing but fractal. Note here that this is concerned with the escape process and 
is different from the fractal associated with classical chaos in a bounded system. In this paper we are concerned with this fractal associated with the escape process. 

\medskip 

There are two kinds of setup to study transient chaos.
\begin{enumerate}
    \item{\it Scattering} \quad 
    One can study literally scattering process when the potential is asymptotically flat. A particle comes from spatial infinity and enter into the scattering region. Then it stays there for a while and exhibits classical chaos. Finally, it escapes from the scattering region and go off to spatial infinity. The scattering angle as a function of impact parameter is a representative physical quantity in this setup.
    For more detailed discussions, see e.g. \cite{scattering-review}.
    

    \item{\it Decay} \quad  Instead of considering an in-coming particle, it is also possible to take the initial condition in the scattering region. Then the particle initially exhibits chaotic motion and then escapes from there after a certain period of time $T$\,, where $T$ is called the time delay function. This escape process may be seen as a kind of instability and $T$ can be seen as lifetime. In comparison to the first case, the asymptotic flatness of the potential is not necessary. When the potential is not bounded from below at spatial infinity, the escape process really means instability. 
    
    In this case, as we will see later, a fractal structure appears in the space of initial conditions for which remain in the scattering region. The singularities in the time delay function $T$ form a Cantor-like set. This fractal is intrinsic to the escape process and different from the long-time behavior case.  
    
\end{enumerate}

We will consider the setup for the second one in a four-hill potential model in Sec.\,\ref{sec:four-hill} and the BFSS matrix model in Sec.\,\ref{sec:matrix}. For the four-hill potential model, both of them can be applied. However, the potential of the BFSS matrix model is not so suitable for the first setup because the potential is very narrow at spatial infinity and there is almost no asymptotic region.


\section{Chaotic instability in a four-hill potential model}\label{sec:four-hill}

We will here focus upon a prototypical model with a four-hill potential. We calculate the time delay function and present the associated fractal structure. 
\subsection{Four-hill potential model}
\begin{figure}[H]
\begin{center}
\begin{minipage}[ht]{0.45\linewidth}
\centering
\includegraphics[width=1\linewidth]{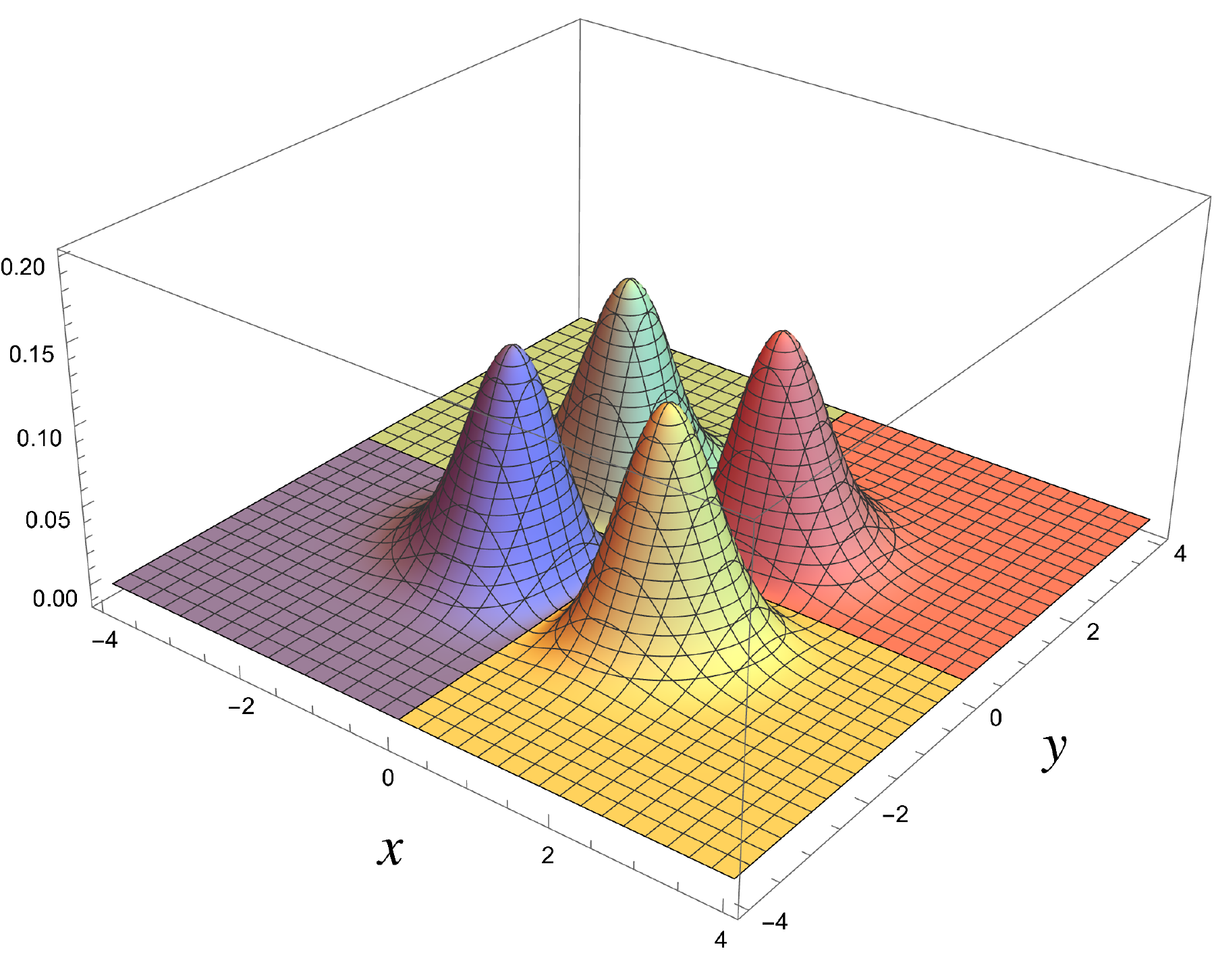}
\caption{\footnotesize The shape of the four-hill potential.
}\label{fig:four-hill potential}
\end{minipage} 
\hspace*{2cm}
\begin{minipage}[ht]{0.35\linewidth}
\centering
\includegraphics[width=1\linewidth]{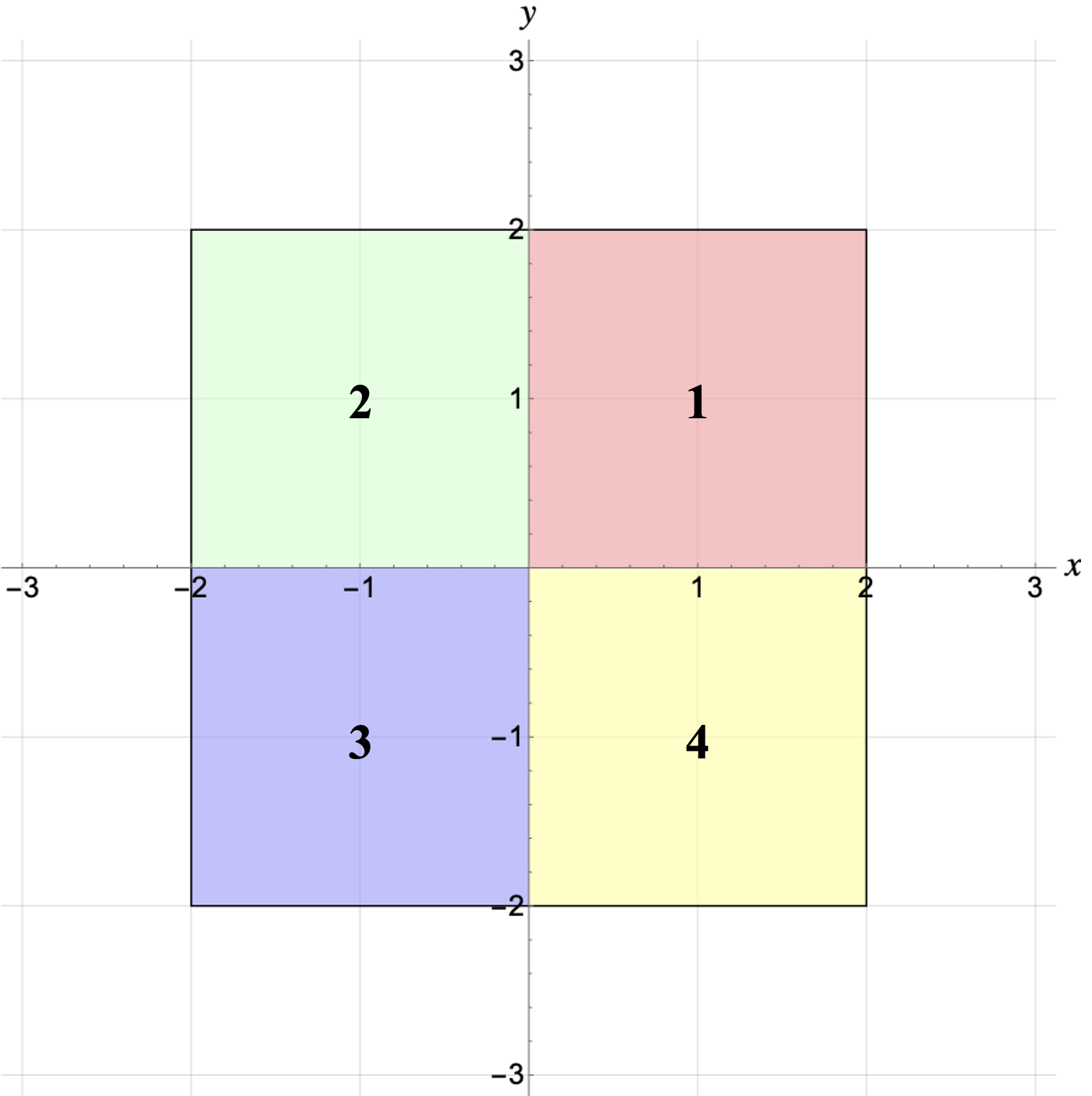}
\caption{\footnotesize The chaotic region $B$\,.
The region $B$ is divided into the four regions.
}\label{fig:chaotic region}
\end{minipage}
\end{center}
\end{figure}

Let us consider the Hamiltonian with a four-hill potential given by
\begin{align}
H=\frac{p_{x}^2}{2} + \frac{p_{y}^2}{2} + x^2y^2\exp(-x^2-y^2)\,,
\end{align}
where the dynamical variables are $x(t)$ and $y(t)$\,, and the conjugate momenta are $p_x(t)$ and $p_y(t)$\,, respectively.
The four-hill potential is flat for $|x|\,,|y|\gg 1$ and has peaks at $(x,y)=(\pm1,\pm1)$\,. The shape of this potential is depicted in Fig.\,\ref{fig:four-hill potential}.
The chaotic region $B$ is defined by $|x|<2$ and $|y|<2$ between the four hills as depicted in Fig.\,\ref{fig:chaotic region}. Since the potential is asymptotically flat, chaotic scattering can be studied in both manners introduced above. Hereafter, we will follow the second manner (decay) because this is close to the analysis of the BFSS matrix model in Sec.\,\ref{sec:matrix}.  

\subsection{Self-similarity in the initial points}

\begin{figure}[htpb]
\begin{minipage}[ht]{0.5\linewidth}
\centering
\includegraphics[width=\linewidth]{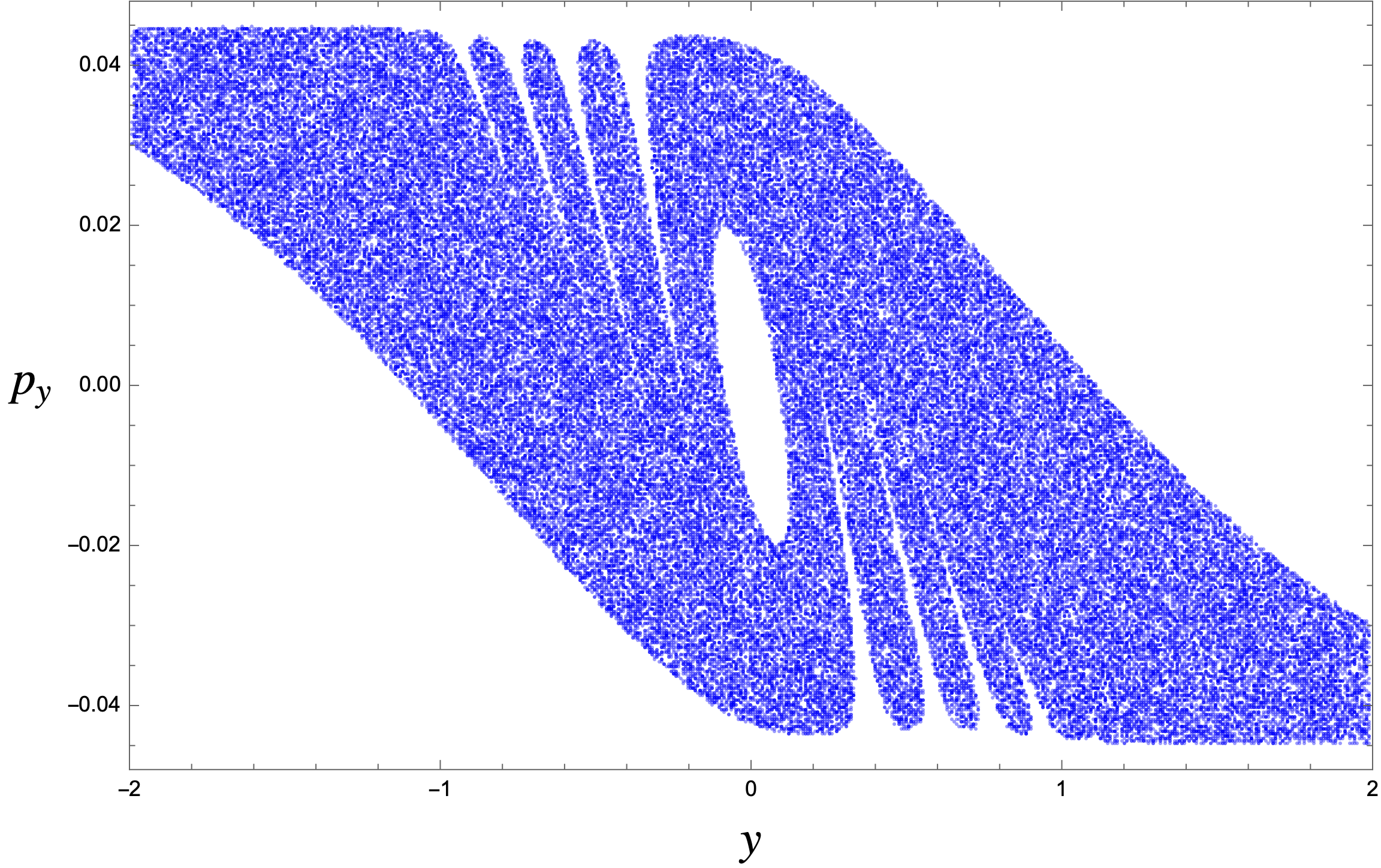}
\end{minipage}
\begin{minipage}[ht]{0.5\linewidth}
\centering
\includegraphics[width=\linewidth]{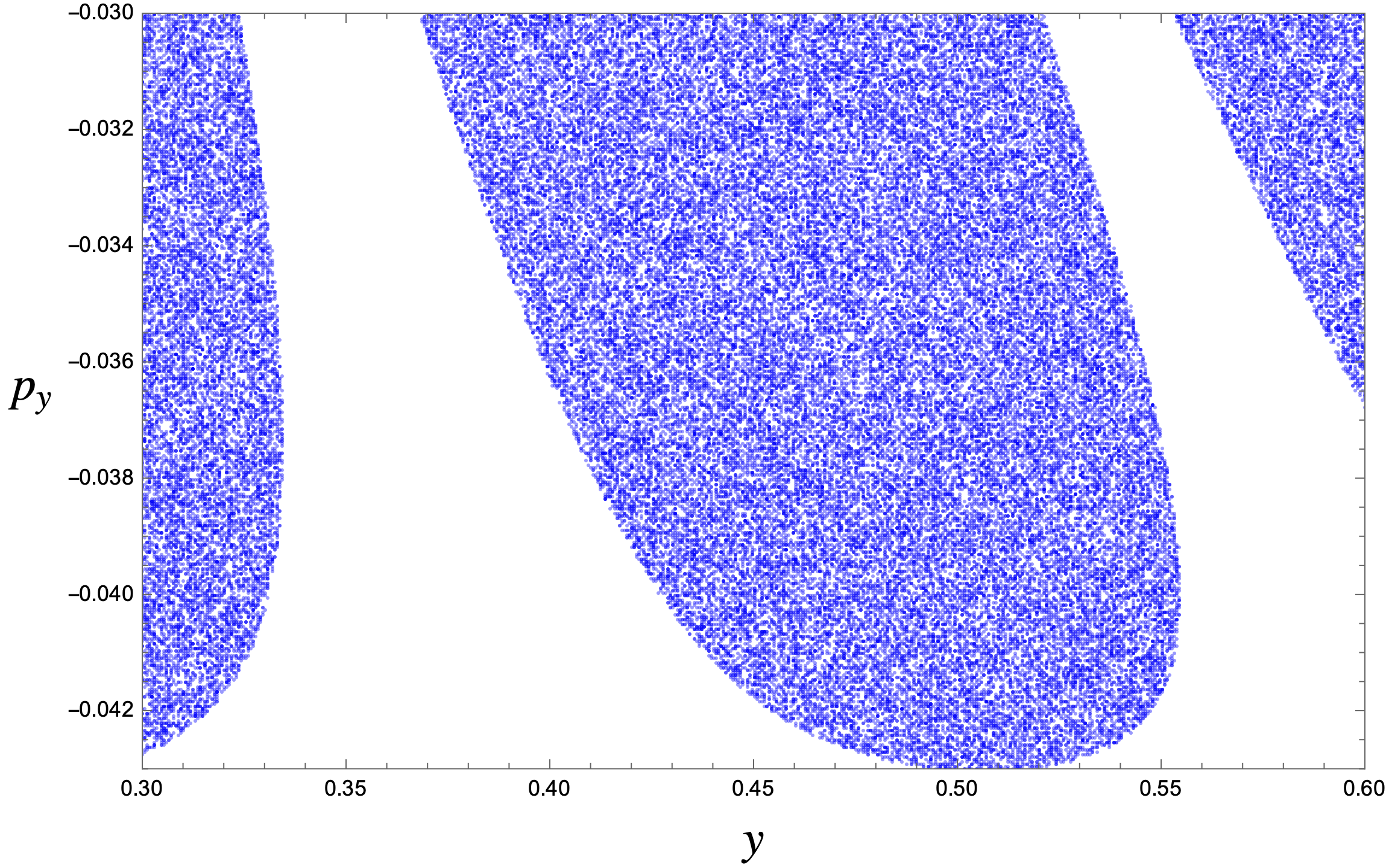}
\end{minipage}
\vspace{-10pt}
\caption*{\footnotesize (a) The initial points which remain at $t=70$}
\vspace{10pt}

\vspace*{0.5cm}

\begin{minipage}[ht]{0.5\linewidth}
\centering
\includegraphics[width=\linewidth]{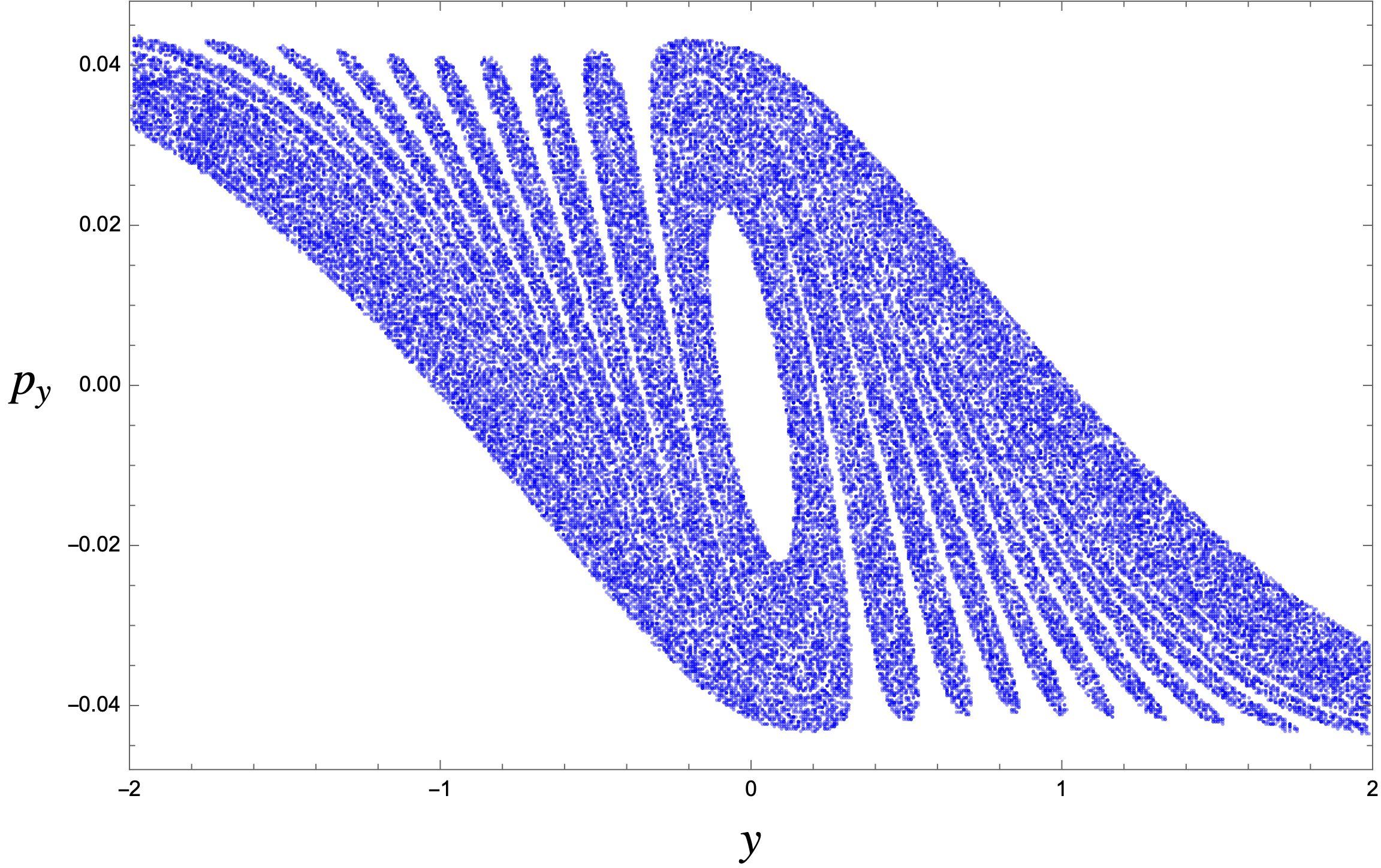}
\end{minipage}
\begin{minipage}[ht]{0.5\linewidth}
\centering
\includegraphics[width=\linewidth]{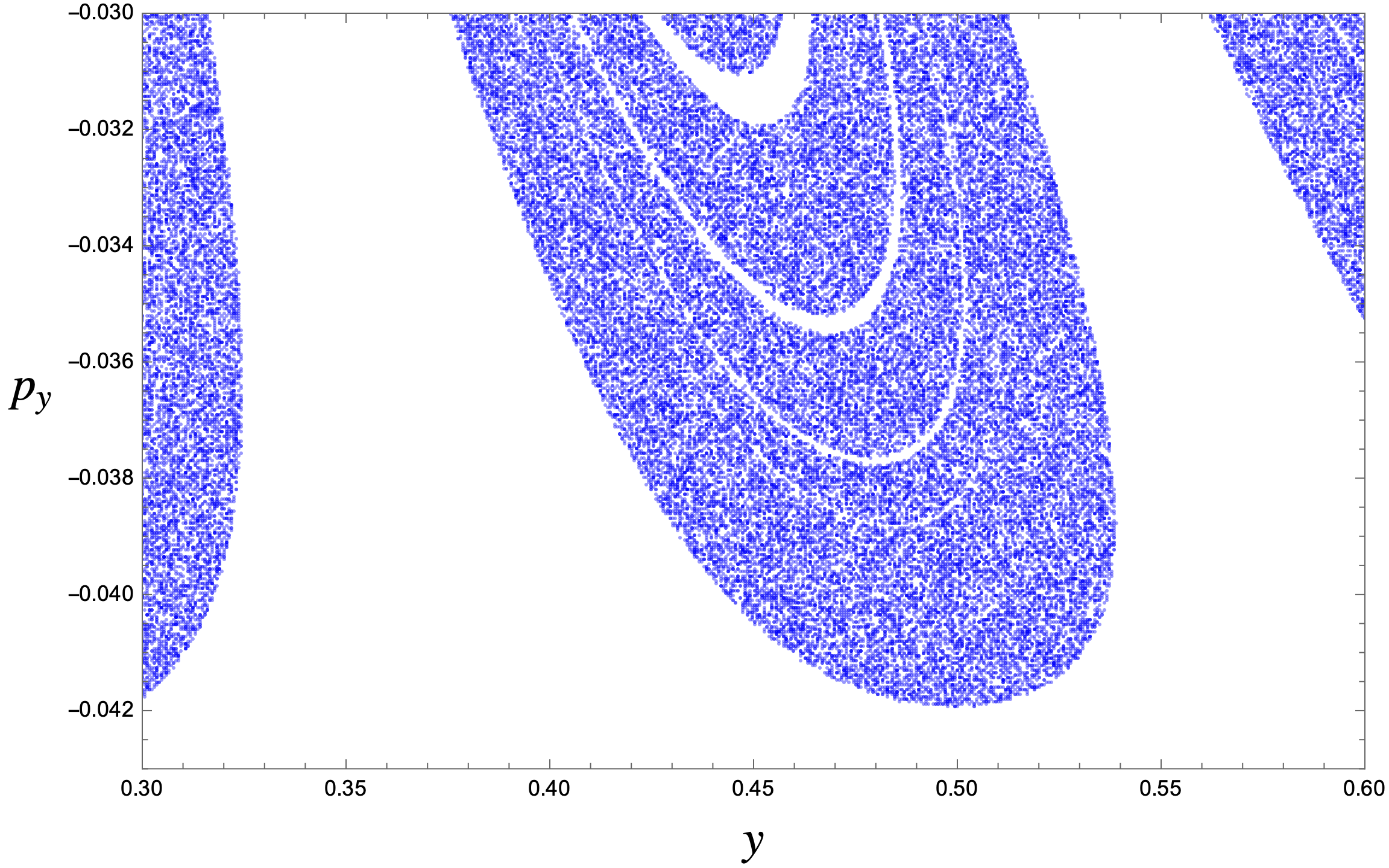}
\end{minipage}
\vspace{-10pt}
\caption*{\footnotesize (b) The initial points which remain at $t=100$}
\vspace{10pt}

\vspace*{0.5cm}

\begin{minipage}[ht]{0.5\linewidth}
\centering
\includegraphics[width=\linewidth]{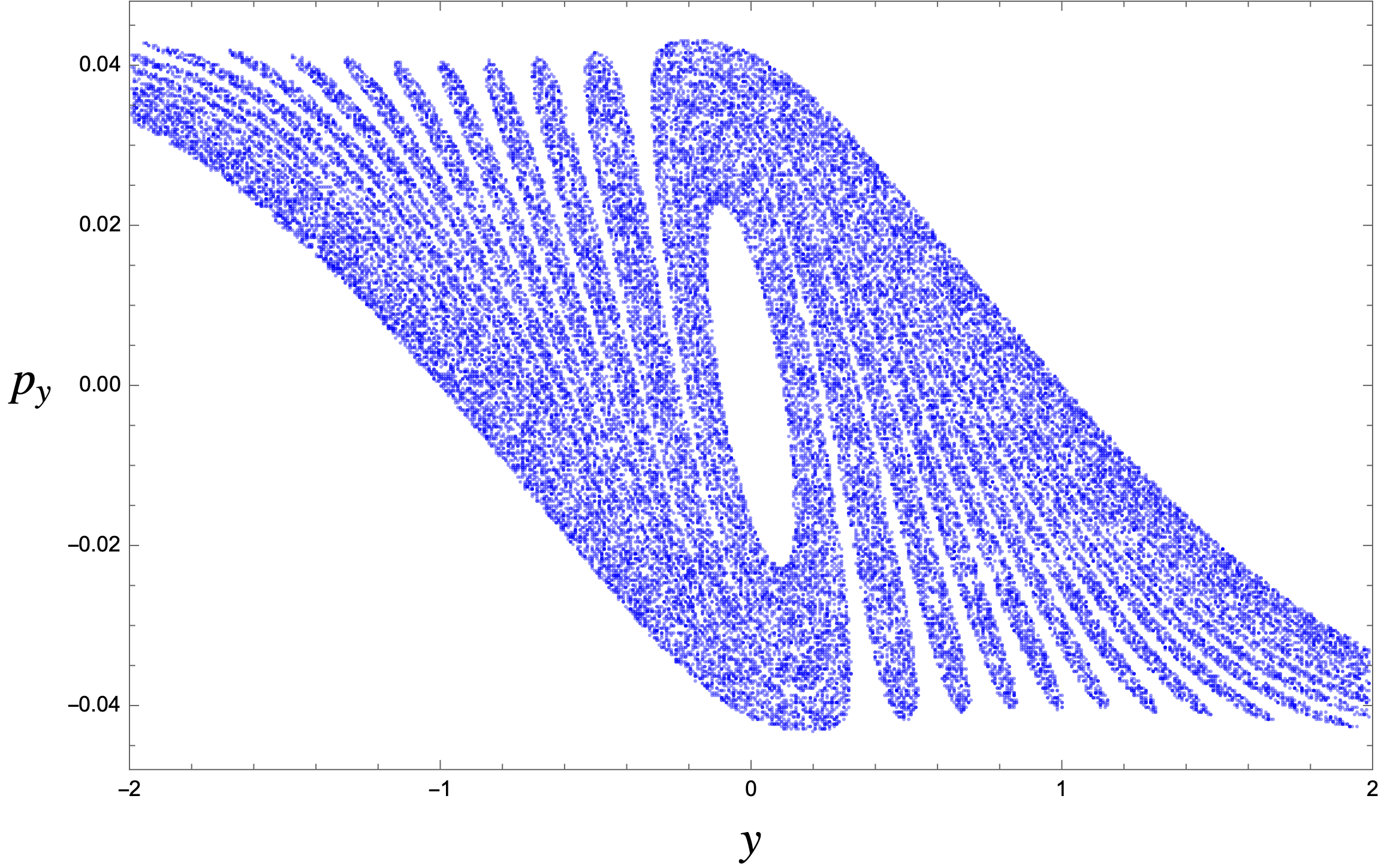}
\end{minipage}
\begin{minipage}[ht]{0.5\linewidth}
\centering
\includegraphics[width=\linewidth]{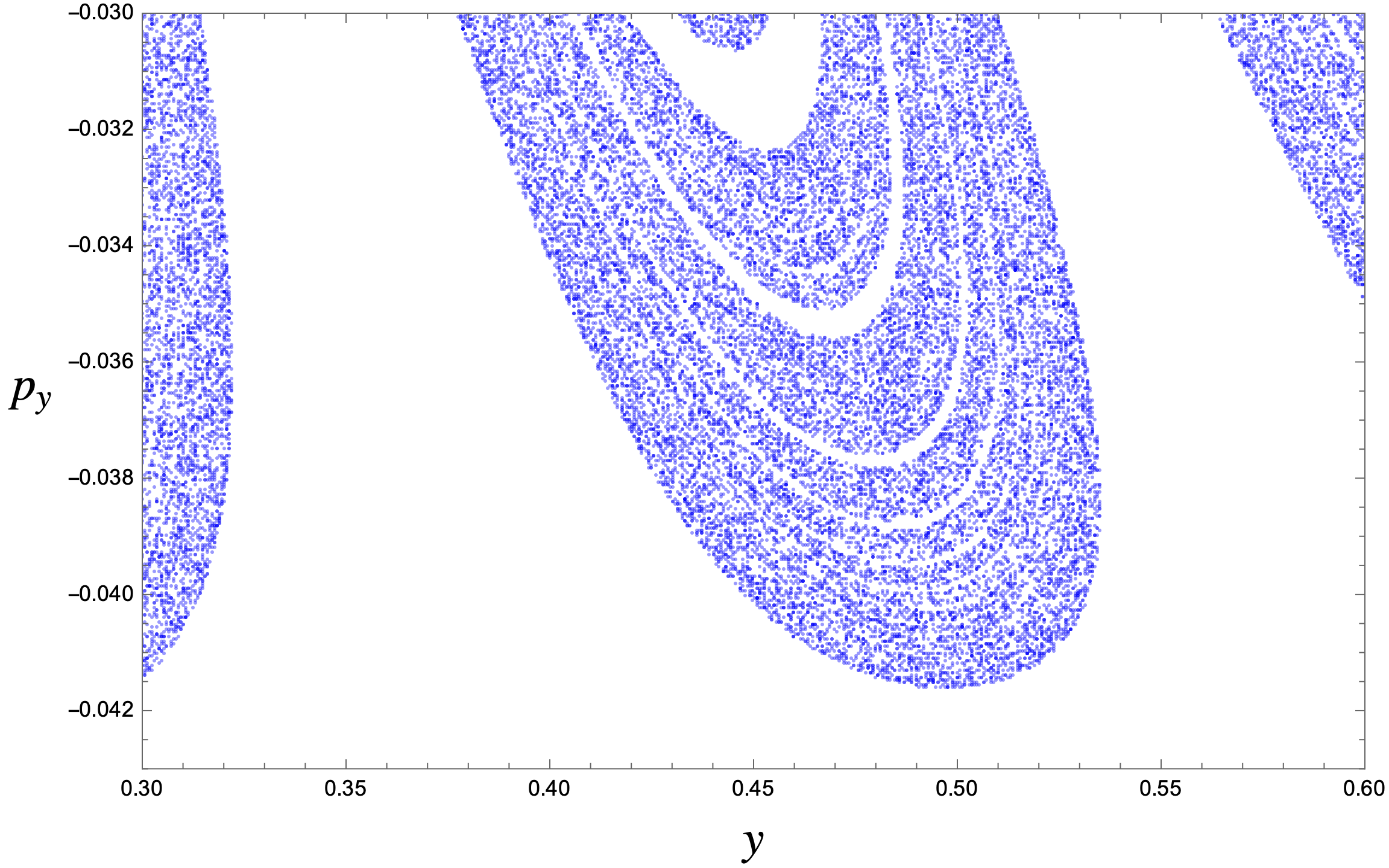}
\end{minipage}
\vspace{-10pt}
\caption*{\footnotesize (c) The initial points which remain at $t=150$}
\caption{\footnotesize The initial conditions which remain in the chaotic region $B$\,. The other parameters are set to $x=0$\,, $p_x>0$ and $E=0.001$\,. The figure on the right is an enlargement of the area of $0.3<y<0.6$ and $-0.043<p_y<-0.03$ in the figure on the left.}\label{fig:four-hill stable}
\end{figure}

Let us consider the $(y,p_y)$-plane of the initial conditions with $x=0$\,,  $p_x>0$ and the energy $E=0.001$ for which trajectories remain in the chaotic region $B$ at a certain time $t$\,. The planes for $t=70$\,, 100 and 150 
are depicted in Fig.\,\ref{fig:four-hill stable}. Remarkably, one can see a self-similarity of gaps in these figures. 
Figure \ref{fig:four-hill stable} (a) illustrates 
the initial conditions for $y$ and $p_y$ for which 
the trajectories remain in the chaotic region $B$ at $t =70$\,. The figure on the right is a magnification of a part of the figure on the left. Similarly, 
Figures \ref{fig:four-hill stable} (b) and (c) illustrate the initial conditions for which the trajectories remain in the chaotic region $B$ at $t=100$ and $t=150$\,, respectively. The density decreases over time because more trajectories escape from the chaotic region $B$\,. Notice in the magnified figures on the right that there is a self-similar structure. At a fixed time $t$\,, smaller structures appear as one area is magnified. The self-similar structure becomes more and more prominent over time. This is nothing but the {\it fractal} structure associated with the chaotic decay.

\subsubsection*{Symbolic dynamics}

A traditional way to interpret a fractal structure behind chaos is to figure out the associated {\it symbolic dynamics}.
A symbolic dynamics is generated by the following manner. 
Let us first divide the chaotic region $B$ into four regions separated by the $x$ and $y$ axes, and each region is labelled by 1, 2, 3 and 4 denoted in Fig.\,\ref{fig:chaotic region}.
Regions 1, 2, 3 and 4 are colored by red, green, blue and yellow, respectively.
Given a trajectory of a particle, one can trace it and record the labels when the particle enters into another region until it leaves the region $B$\,.
The trajectory is then represented by a sequence of the labels such as $1\rightarrow 2 \rightarrow 1 \rightarrow 4 \rightarrow 3\dots$ ($12143\dots$ for short).

\medskip

Consider a trajectory starting from a region, say 1.
Depending on the initial condition, the particle can return to the region 1 through 121, 12321, 141 and so on.
However, one can notice that there are also trajectories where the particle escapes between the hills.
When plotting the initial points on the Poincar\'e section which remain in $B$\,, these escaping trajectories appear as gaps.
Once the particle comes back to the region 1, the same procedure is repeated for infinite times.
The Poincar\'e map for this system is thus given by a Smale horseshoe map, which is defined as infinite iterations of a map $f$ from the $(y,p_y)$-plane into itself with stretching and folding.

\medskip

The horseshoe construction indicates that the number of symbolic dynamics grows exponentially.
It is the origin of infinitely self-similar structure, fractal.
Indeed, the number of distinct trajectories is uncountable. This property distinguishes transient chaos from the other erratic behaviors.

\medskip

The horseshoe interpretation is valid for both chaos in a bounded space and transient chaos. In the latter case, although almost all trajectories are escaping, specific trajectories still remain in $B$ after infinite time.
The simplest case is with the initial condition $x=y>0$ and $p_x=p_y<0$\,.
In this case, the symbolic dynamics is described by $131313\dots$\,, and never ends.

\subsection{Time delay function}

\begin{figure}[htpb]
\vspace*{0.5cm}
\begin{minipage}[ht]{0.5\linewidth}
\centering
\includegraphics[scale=0.4]{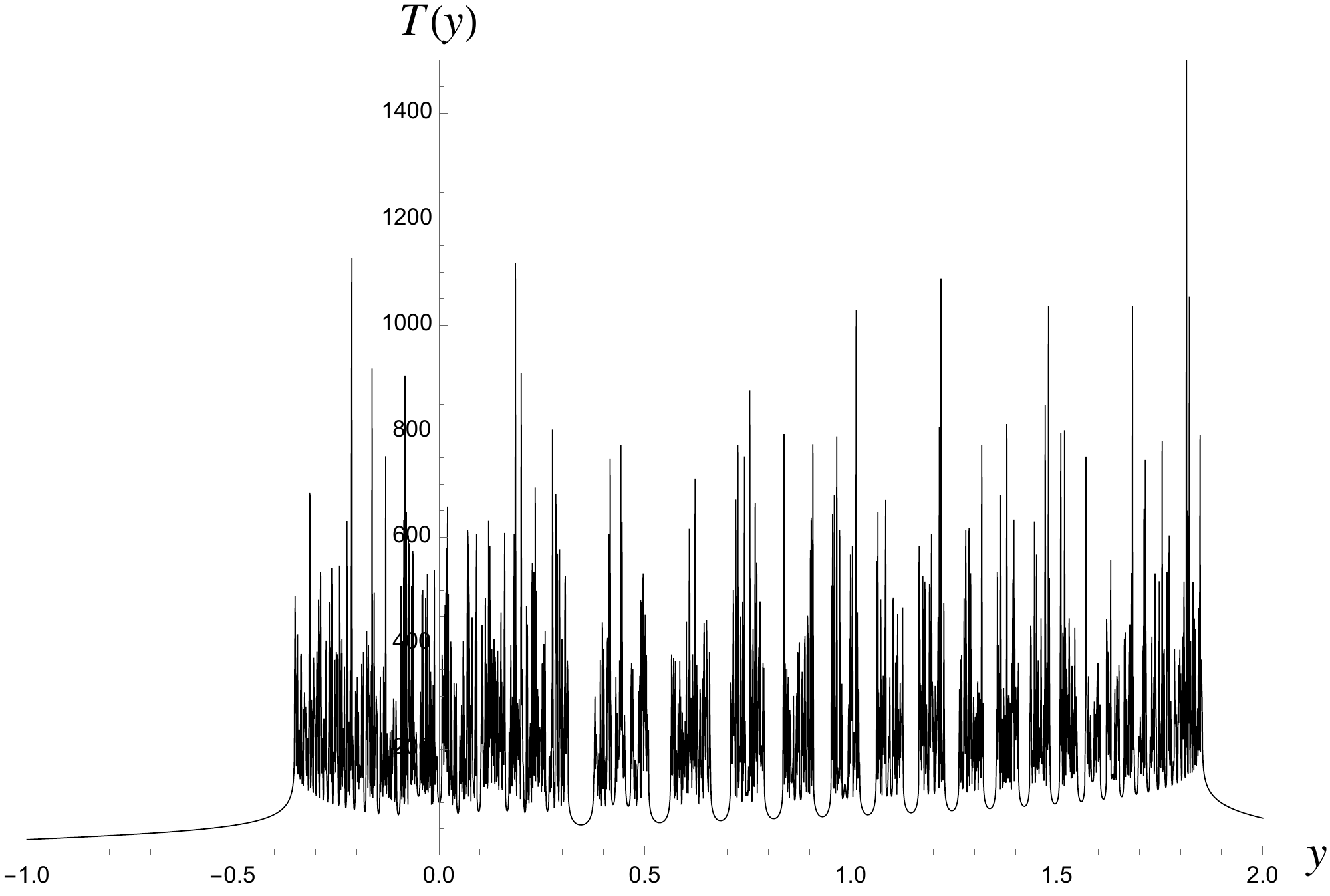}
\end{minipage}
\begin{minipage}[ht]{0.5\linewidth}
\centering
\includegraphics[scale=0.4]{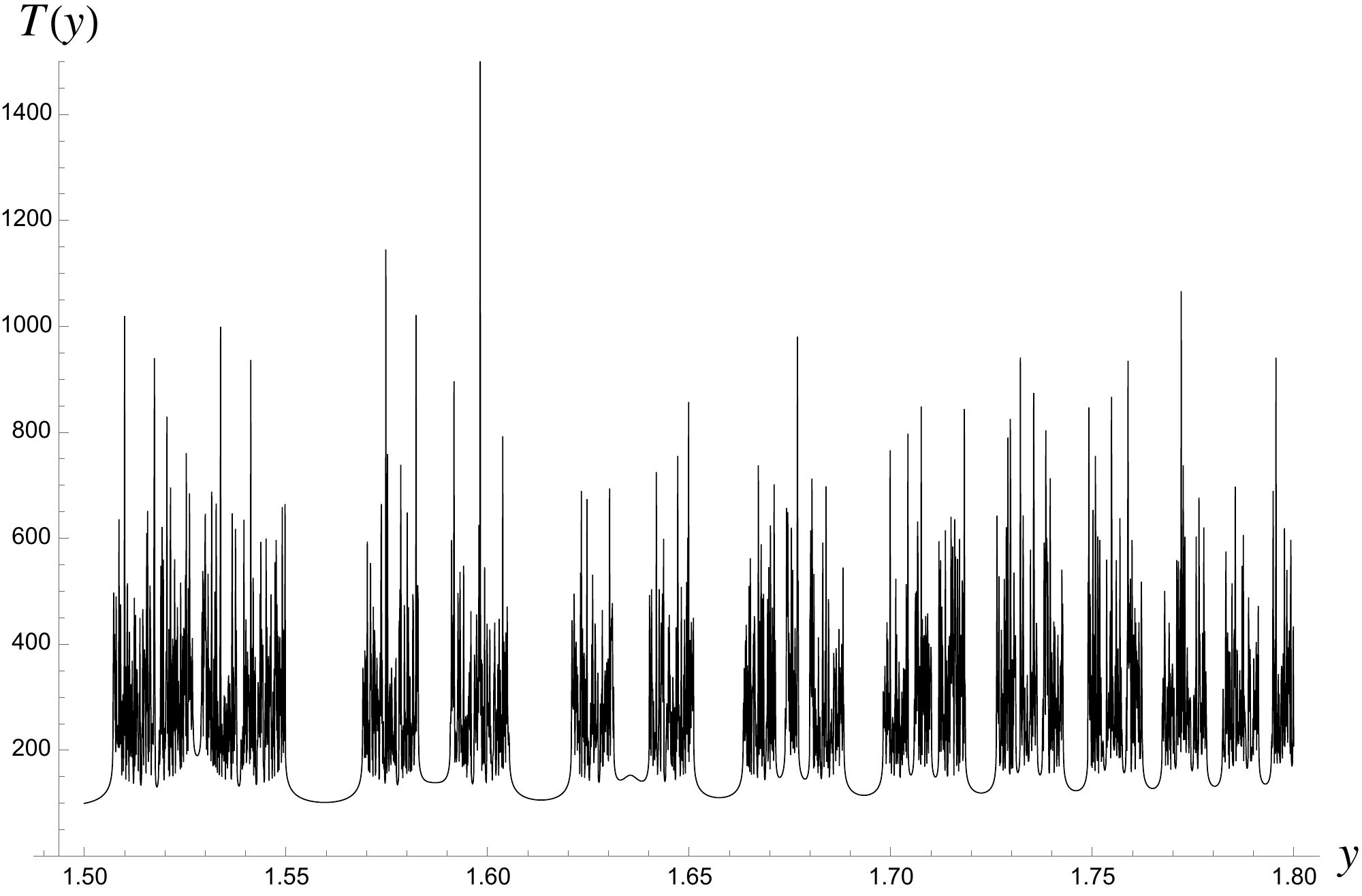}
\end{minipage}
\caption{\footnotesize The escape time $T$ with the initial condition $x=0$\,, $p_y=-0.03$ and $E=0.001$\,.
The right figure zooms in on the area of $1.5<y<1.8$ in the left figure.
}\label{fig:time four-hill}
\end{figure}

Another feature of chaotic scattering can be seen in the behavior of the time delay function $T$\,.
Figure \ref{fig:time four-hill} shows the time $T$ to escape from the chaotic region $B$ with respect to the initial condition.
The initial condition is distributed for $x=0$ in $B$\,, with $p_{y}=-0.03$ and $E=0.001$\,. Namely, only $y$ is an argument of $T$\,. The singular points corresponds to long-lived trajectories. The smooth intervals correspond to the gaps in Fig.\,\ref{fig:four-hill stable}. 
The figure on the right is a zoom-up of an area of the figure on the left. Even after magnifying the area, similar structure of smooth intervals appear. This self-similarity continues infinitely. Thus one can say that the singular points form a Cantor-like set\footnote{In the analogy with the Cantor ternary set, the smooth interval corresponds to the middle 1/3 interval to be truncated.} $\cC\subset[-2,2]$ though it is quite difficult to identify the dynamical map for this set concretely. In fact, the fractal dimension $d$ (or box-counting dimension) of this set $d~(0<d<1)$ can be estimated as we will discuss below. 

\medskip

In the following, let us estimate the fractal dimension of $\cC$ by means of the {\it uncertainty fraction} $f(\epsilon)$\,. 
Choosing a random point $x_0\in[-2,2]$\,, we say that $x_0$ is $\epsilon$-uncertain if $|T(x_0+\epsilon)-T(x_0)|>h$\,, where 
the positive number $h$ is taken to be a typical escaping time, say 100.
We take random reference points $x_0$\,, and tell whether these are $\epsilon$-uncertain or not.
The uncertainty fraction $f(\epsilon)$ is obtained by dividing the number of uncertain points by the total number of chosen random points.
We then define the uncertainty dimension by
\begin{align}
d_{\rm u}:=1 - \lim_{\epsilon\to +0}\frac{\log f(\epsilon)}{\log\epsilon}\,.
\end{align}

\medskip 

The box-counting dimension $d_{\rm box}$ is often adopted as a fractal dimension.
In typical dynamical systems, 
$d_{\rm box}$ is shown to be equal to the uncertainty dimension $d_{\rm u}$\cite{dimension}\,.
To understand the relation between these dimensions, 
suppose $f(\epsilon)$ equals to the probability such that a singularity is included in $x\in[x_0,x_0+\epsilon]$\,. 
Under this assumption,
the box-counting dimension is given by
\begin{align}
d_{\rm box} = \lim_{\epsilon\to +0}\frac{\log(4/\epsilon\cdot f(\epsilon))}{\log(4/\epsilon)}
=1+\lim_{\epsilon\to +0}\frac{\log f(\epsilon)}{\log(4/\epsilon)}
= d_{\rm u}\,.
\end{align}
In this paper, we refer to the uncertain dimension $d_{\rm u}$ as the fractal dimension $d$\,, i.e.,
\begin{eqnarray}
 d 
 = d_{\rm u}\,. 
\end{eqnarray}

\medskip

\begin{figure}
\centering
\includegraphics[width=0.6\linewidth]{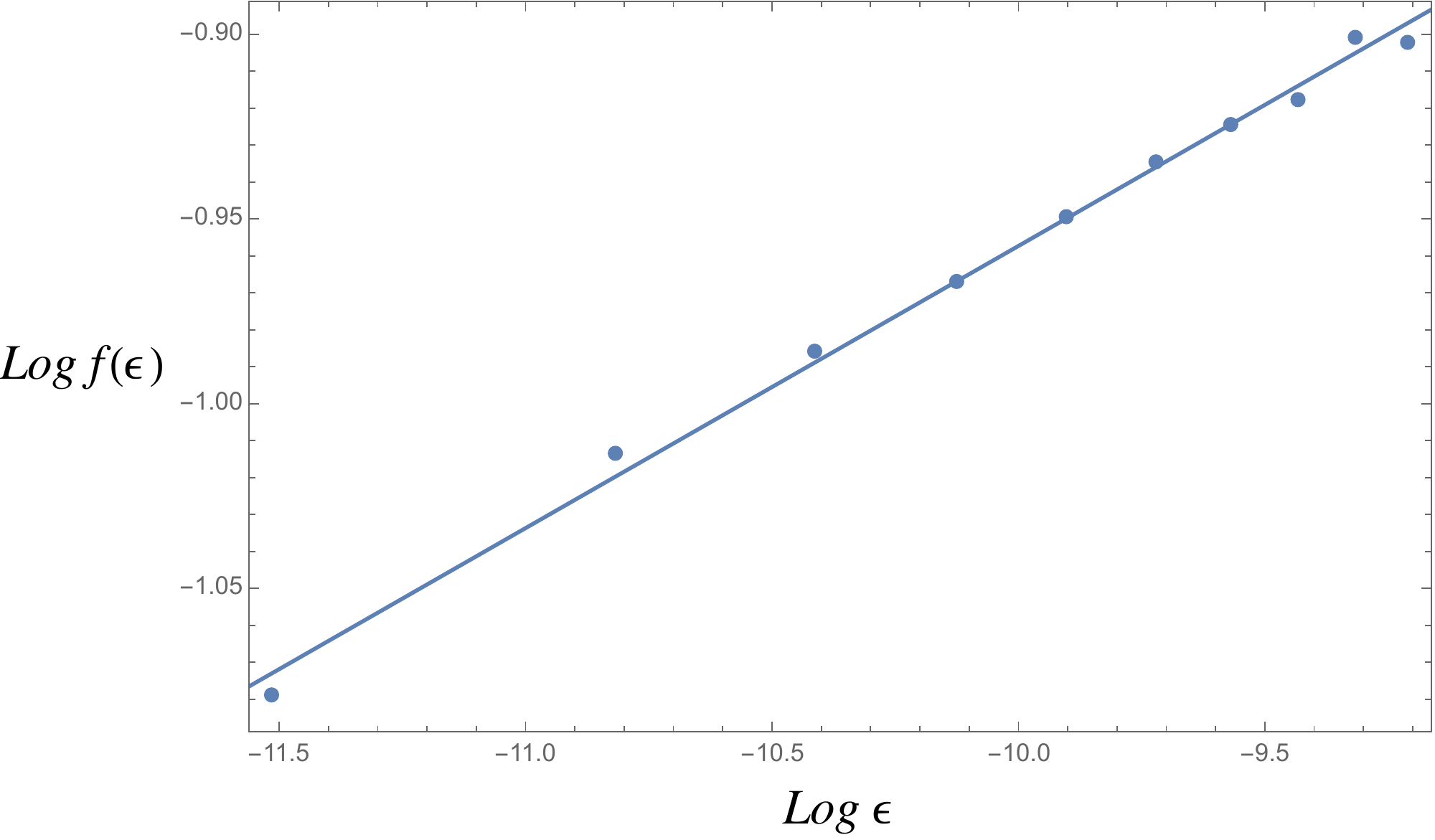}
\caption{\footnotesize The uncertain fraction with $h=50$\,.
The axes are given by the natural logarithms.}\label{fig:four-hill time log}
\vspace*{0.5cm}
\end{figure}

The uncertain dimension can be obtained by fitting the plot of $\log f(\epsilon)$ versus $\log \epsilon$ as shown in Fig.\,\ref{fig:four-hill time log}. For each value of $\epsilon$\,, $10^4$ points are taken as random reference points $x_0$\,. From this figure, one can read off the following relation:  
\begin{align}
    \log f(\epsilon) =(0.0764\pm 0.0019) \log \epsilon + (-0.194\pm 0.019)\,. 
\end{align}
Thus the fractal dimension $d$ for this initial condition is given by $d=0.924\pm 0.002$\,.

\section{Chaotic instability in the BFSS matrix model}\label{sec:matrix}
In this section, we shall discuss chaotic scattering in the BFSS matrix model.
The BFSS matrix model is a one-dimensional $U(N)$ gauge theory with adjoint matters. The bosonic part of the classical action is given by
\begin{align}
S=&\,\int dt\, \tr( \frac{1}{2}\sum_{r=1}^{9}(DX^{r})^2 + \frac{1}{4}\sum_{r,s=1}^{9} \big[ X^{r},X^{s}\big]^2 )\,,
\label{BFSS action}
\\
&  DX^{r}:=\, \frac{d}{dt}X^{r} - i [A,X^{r}]\,,
\end{align}
where $X^{r}(\tau)$ are $N\times N$ hermitian matrix-valued variables.
The equation of motion for a gauge field $A$ leads to the Gauss-law constraint
\begin{align}
0=\sum_{r=1}^{9}\left[ X^{r},DX^{r}\right]\,.\label{Gauss law}
\end{align}
By taking a variation with respect to $X^r$\,, the equations of motion are obtained as
\begin{align}
0=\frac{d^2}{dt^2}X^{r} + \sum_{s=1}^{9}\big[ X^{s}, \big[X^{s},X^r\big]\big]\,,
\label{matrix eom}
\end{align}
where we have taken a static gauge $A=0$ and hereafter will work in this gauge.
Note that the classical action (\ref{BFSS action}) has flat directions along the Abelian subalgebra. 
Since this property allows unbounded motions, one may discuss chaotic scattering in this system. 
In the following, we consider the $N=2$ case for simplicity. 

\medskip

Next,  let us reduce the equations of motion 
by supposing a specific ansatz: 
\begin{align}
X^1(t) &=x(t)\frac{\sigma^1}{2}\,,\quad X^2(t)=y(t)\frac{\sigma^2}{2}\,,\quad
X^3(t)=z(t)\frac{\sigma^3}{2}\,, \label{sphere ansatz}\\ 
X^{s}(t) &=0 \qquad (s=4,\dots,9)\,. \notag
\end{align}
Here $\sigma^1$\,, $\sigma^2$ and $\sigma^3$ are the Pauli matrices and $x(t)$\,, $y(t)$ and $z(t)$ are functions to be determined.
The Gauss Law constraint (\ref{Gauss law}) is automatically satisfied under the ansatz (\ref{sphere ansatz}).
In light of \cite{Kabat:1997im}, 
this configuration can be interpreted as a pulsating membrane.
The equations of motion (\ref{matrix eom}) are then reduced to
\begin{align}\begin{split}
0=&\,\frac{d^2}{d\tau^2}x(t)+x(t)\big(y(t)^2+z(t)^2\big)\,,\\
0=&\,\frac{d^2}{d\tau^2}y(t)+y(t)\big(z(t)^2+x(t)^2\big)\,,\\
0=&\,\frac{d^2}{d\tau^2}z(t)+z(t)\big(x(t)^2+y(t)^2\big)\,.
\end{split}\label{EOM xyz}
\end{align}

\medskip 

Here, let us impose a further constraint $z(t)=0$\,. This does not spoil down the Gauss law constraint (\ref{Gauss law}). 
From the point of view of membrane dynamics, this constraint means that the diameter for the $X^3$-direction is set to zero.
Under this constraint, the equations of motion (\ref{EOM xyz}) are further reduced to 
\begin{align}
0=\frac{d^2}{dt^2}x(t)+x(t)y(t)^2\,,\qquad
0=\frac{d^2}{dt^2}y(t)+x(t)^2y(t)\,. 
\end{align}
Note that these equations of motion are derived also from the Hamiltonian
\begin{align}
H=\frac{p_{x}^2}{2} + \frac{p_{y}^2}{2} + \frac{x^2y^2}{2}\,.
\label{xy potential}
\end{align}
This system was also considered in \cite{Matinyan:1981dj,Arefeva:1997oyf}. 
Note here that this system (\ref{xy potential}) 
can also be derived from a four-dimensional classical Yang-Mills theories as shown in \cite{Matinyan:1981dj}. Hence the results we will obtain below hold for the four-dimensional Yang-Mills theory beyond the BFSS matrix model. 

\medskip

\begin{figure}[htpb]
\begin{center}
\begin{minipage}[ht]{0.3\linewidth}
\centering
\includegraphics[width=\linewidth]{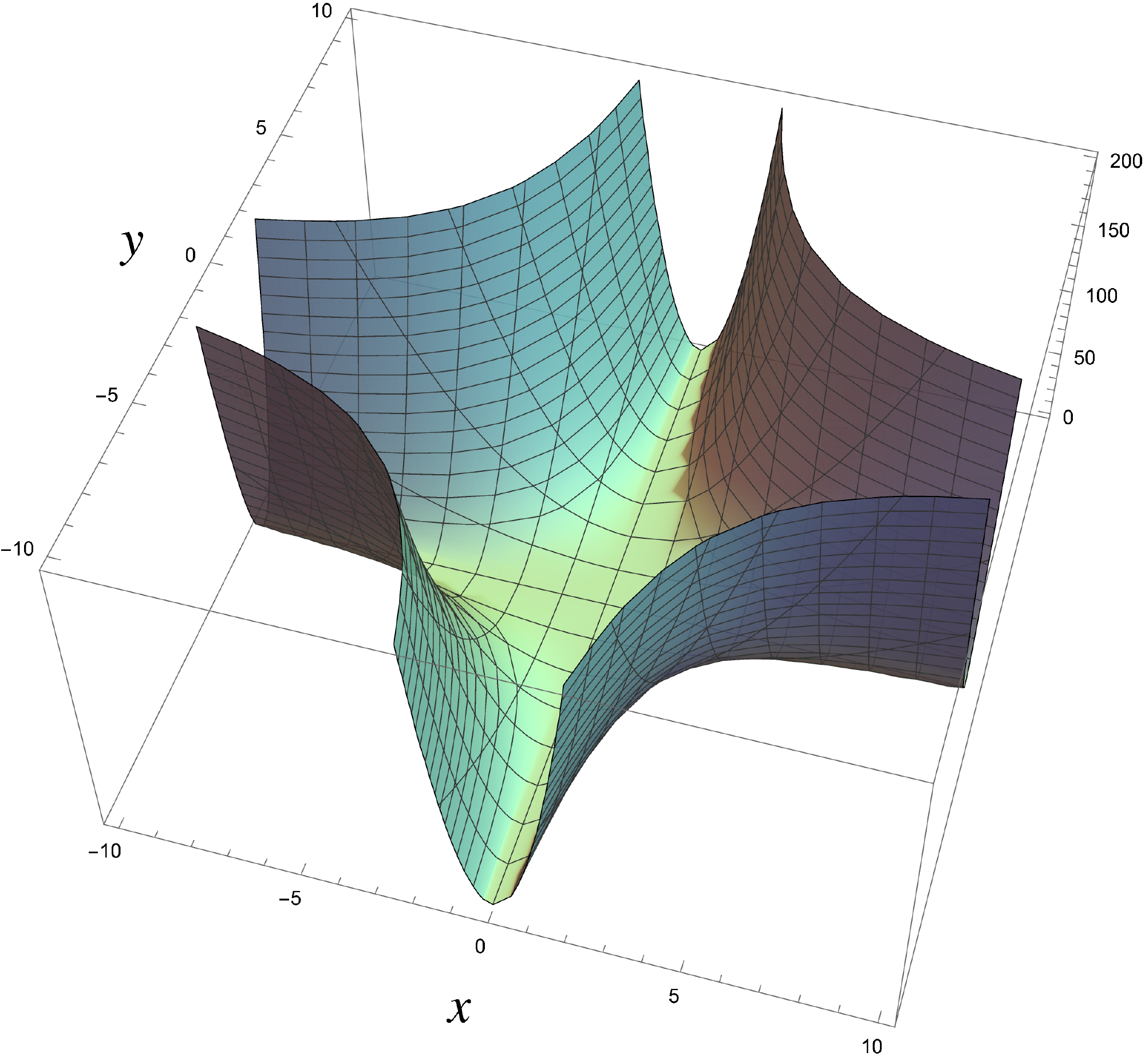}
\caption{\footnotesize The shape of the potential $x^2y^2/2$\,.}\label{fig:matrix potential}
\end{minipage}
\hspace*{1cm}
\begin{minipage}[ht]{0.5\linewidth}
\centering
\includegraphics[width=\linewidth]{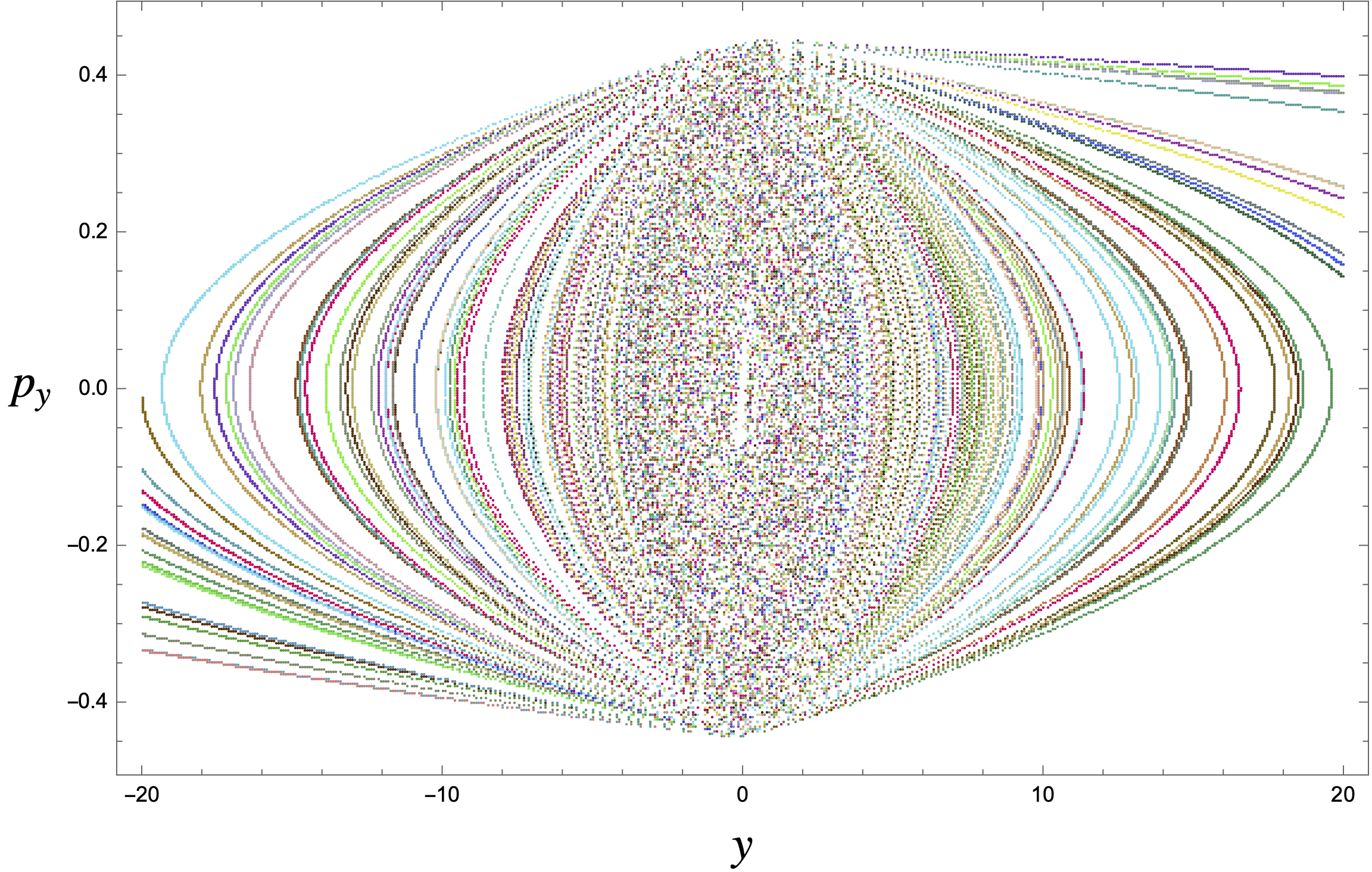}
\caption{\footnotesize Poincar\'e sections for $E=0.1$.
Each color represents a different trajectory.
}\label{matrix Poincare}
\end{minipage}
\end{center}
\end{figure}

The potential in (\ref{xy potential}) is zero on the $x$- and $y$-axes (see Fig.\,\ref{fig:matrix potential}) and hence the diameter in either direction can be expanded to any width in a finite amount of time. Therefore, one may consider the membrane decay as a process of chaotic scattering and discuss the associated fractal structure as in the four-hill potential case. 

\medskip

Following the standard method, we can compute Poincar\'e sections as shown in Fig.\,\ref{matrix Poincare}.
Calculations are stopped if orbits reach $|y|=20$ so as to avoid numerical errors.
When $E=0.1$\,, motions are no longer chaotic once the point leaves the region around $|y|\leq 4$\,.
Therefore, we take the chaotic region $B$ as the box $|x|<5$\,, $|y|<5$\,.
Outside the region $B$\,, points go far away along the ``paths'' for $x=0$ or $y=0$ while oscillating in the ``valleys'' \footnote{For an alternative approach to define the lifetime, see an intriguing work \cite{Berenstein}.}.

\medskip

\begin{figure}[htbp]
\centering
\begin{minipage}[ht]{0.47\linewidth}
\centering
\hspace*{-1.5cm}
\includegraphics[width=\linewidth]{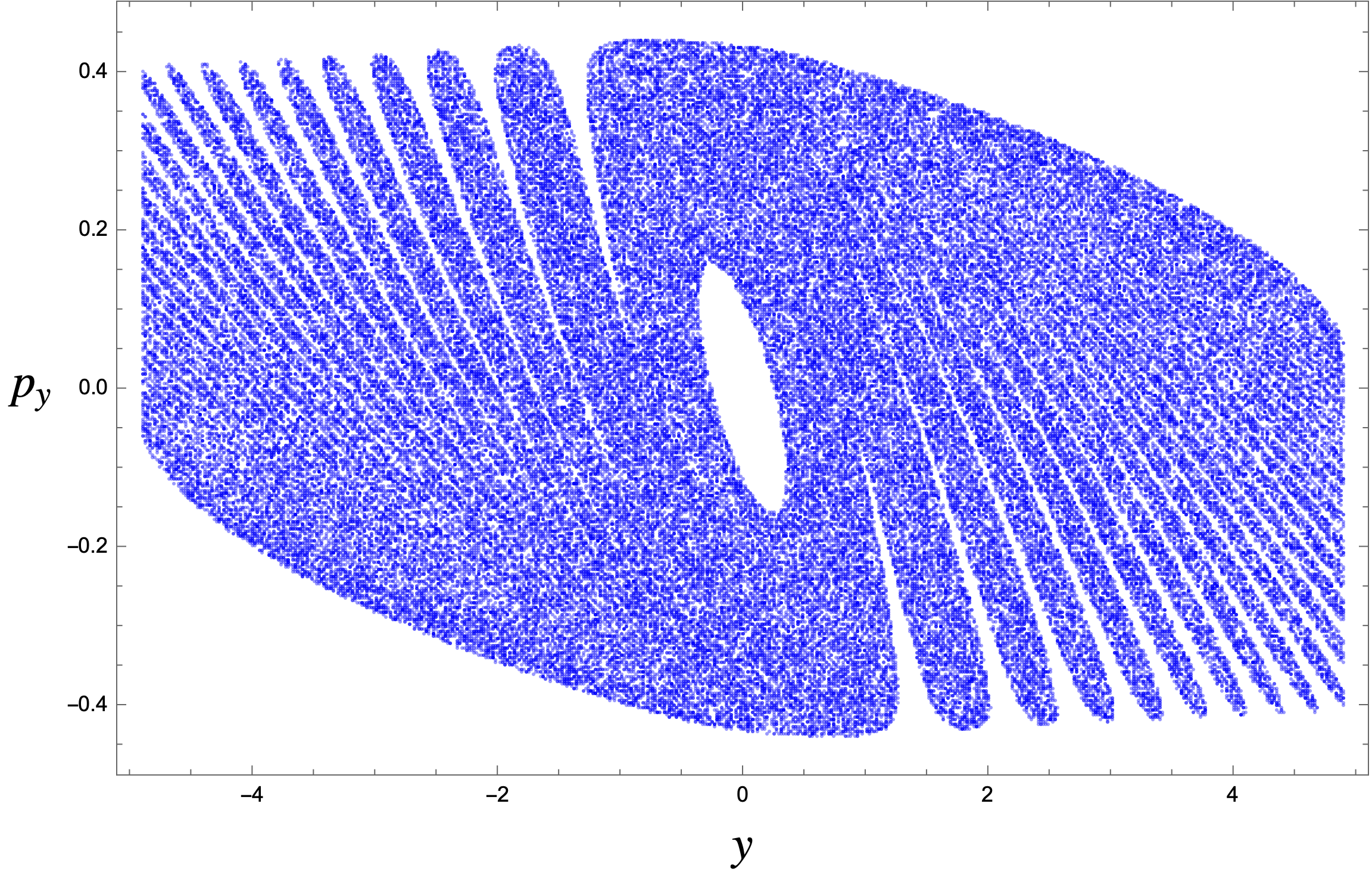}
\end{minipage}
\begin{minipage}[ht]{0.47\linewidth}
\centering
\hspace*{0.3cm}
\includegraphics[width=\linewidth]{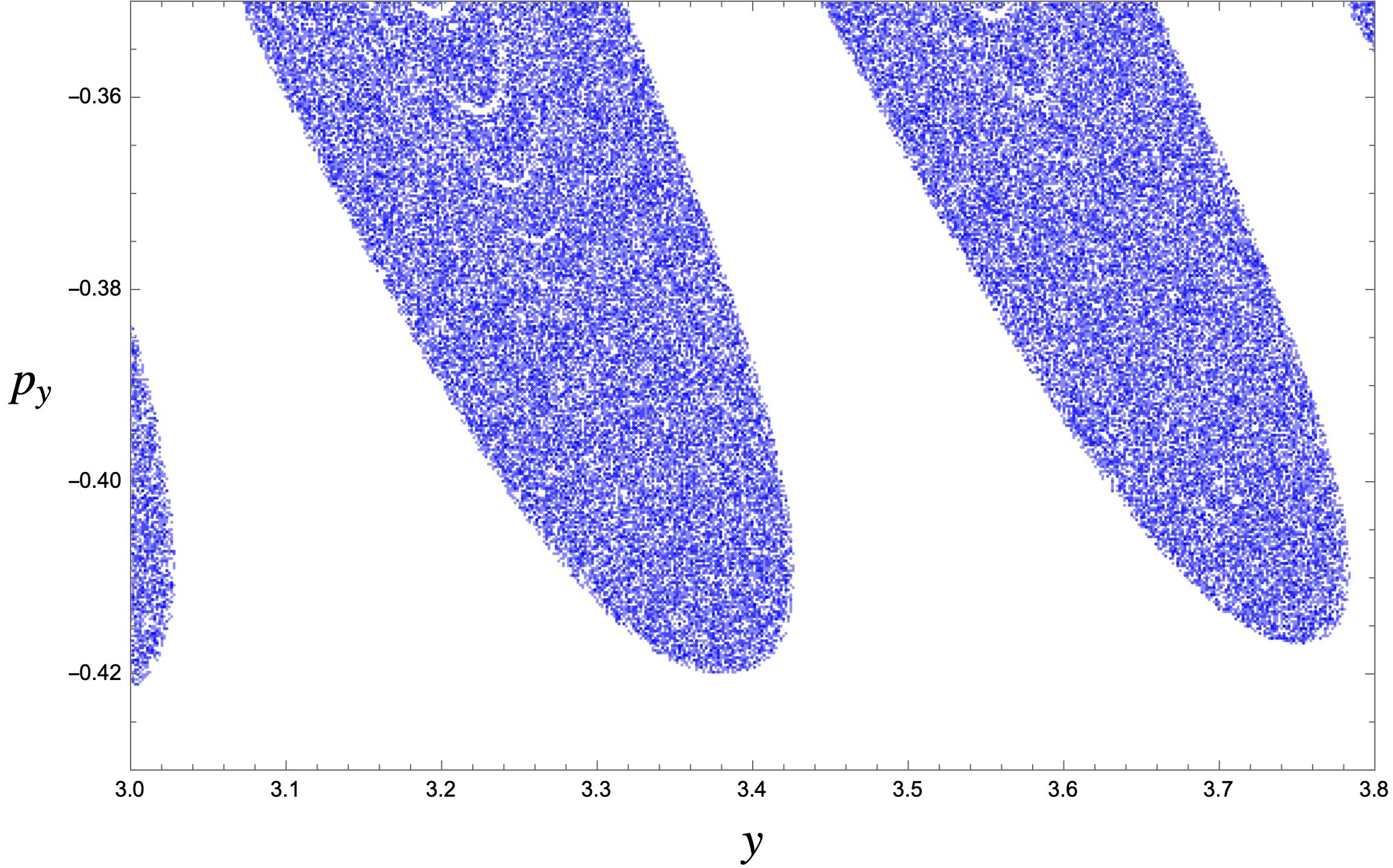}
\end{minipage}
\vspace{-10pt}
\caption*{\footnotesize (a) The initial points which remain at $t=40$}
\vspace{10pt}

\begin{minipage}[ht]{0.47\linewidth}
\centering
\hspace*{-1.5cm}
\includegraphics[width=\linewidth]{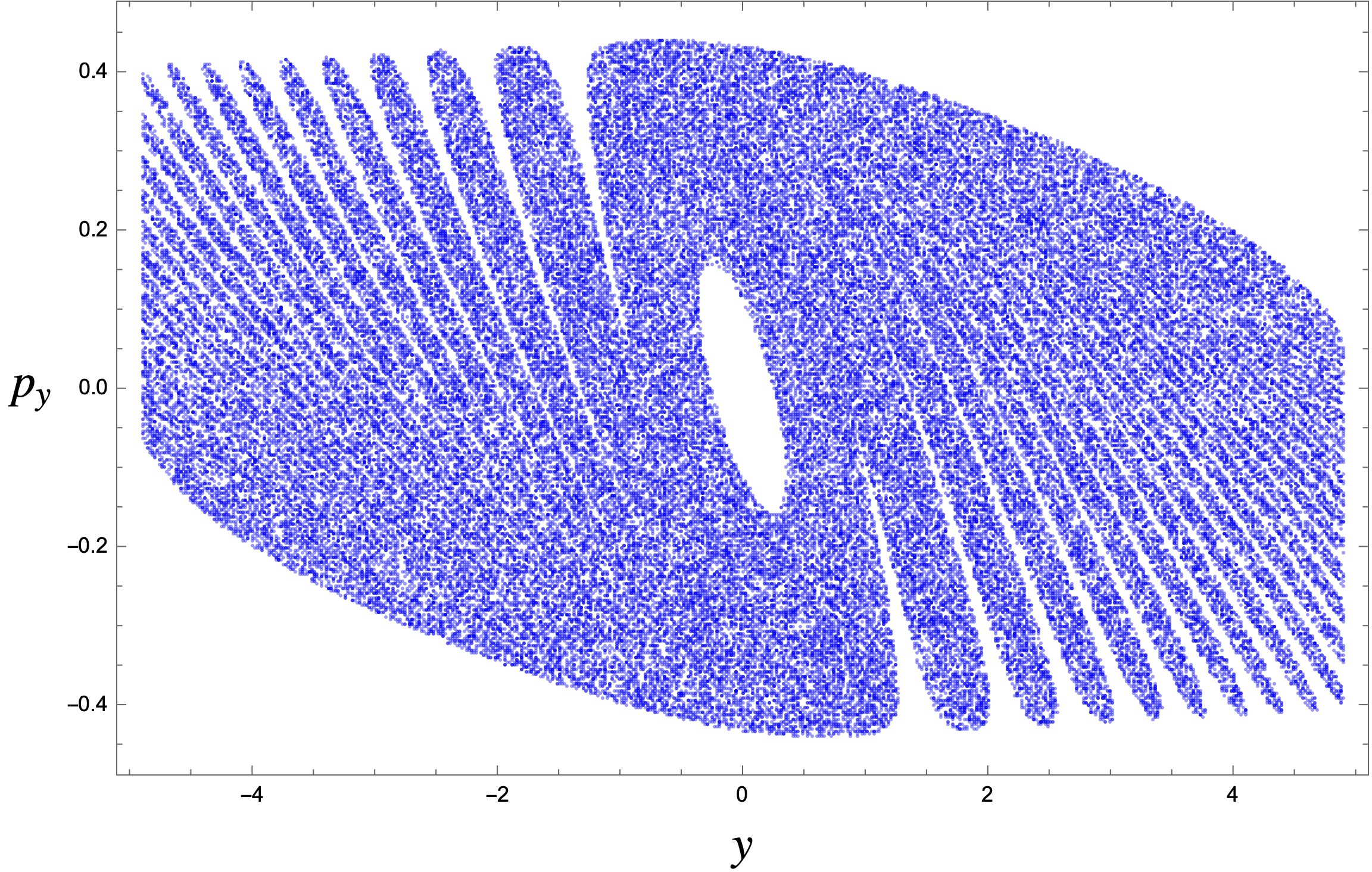}
\end{minipage}
\begin{minipage}[ht]{0.47\linewidth}
\centering
\hspace*{0.3cm}
\includegraphics[width=\linewidth]{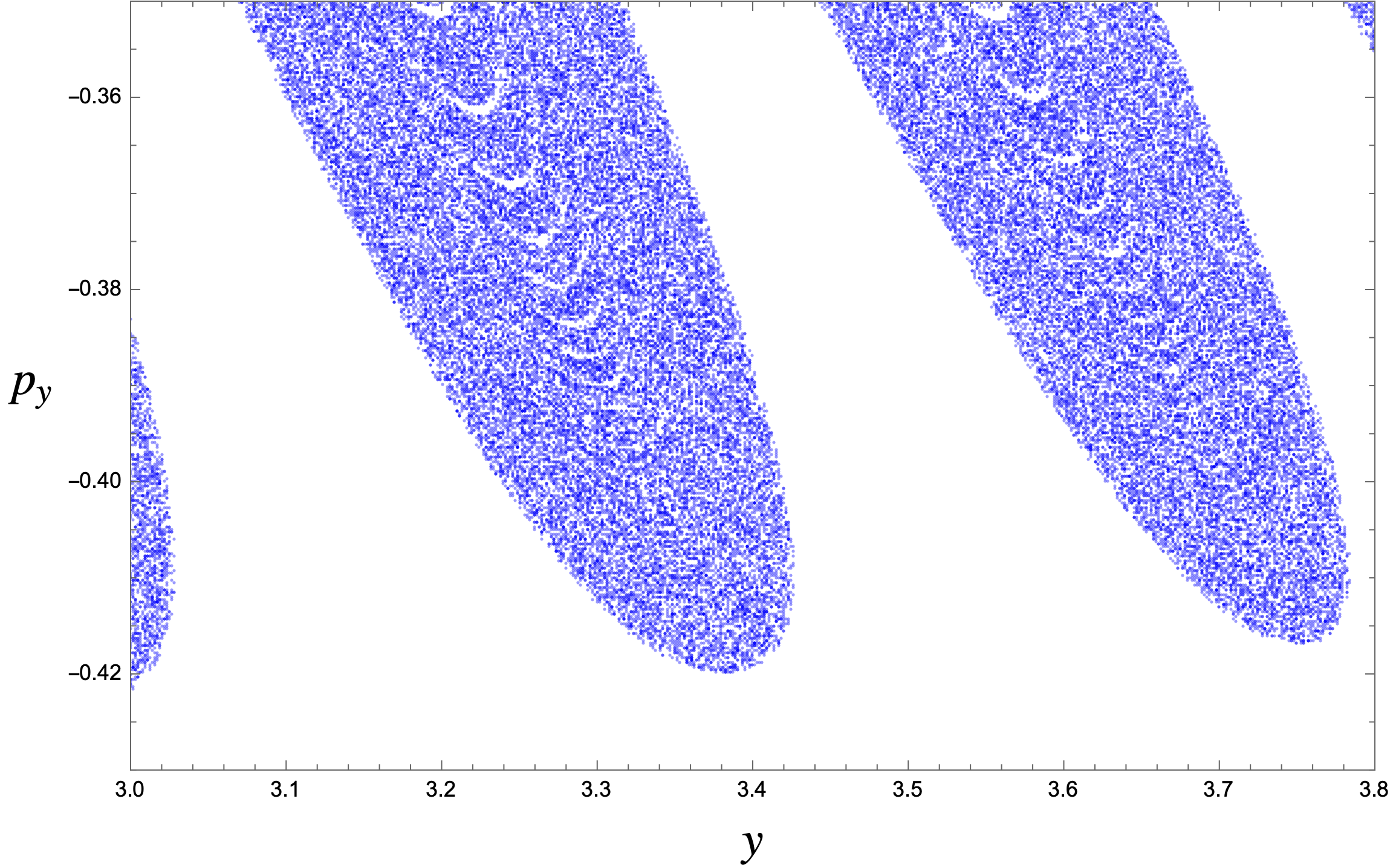}
\end{minipage}
\vspace{-10pt}
\caption*{\footnotesize (b) The initial points which remain at $t=70$}
\vspace{10pt}
\caption{\footnotesize Initial conditions which remain in the chaotic region $B$\,.
The figure on the right is a zoomed-in view of the area of $3.0<y<3.8$\,, $-0.043<p_y<-0.35$ in the figure on the left.}\label{matrix stable}
\end{figure}

We can calculate 
remaining trajectories and time delay function by the same procedure as in Sec.\,\ref{sec:four-hill}.
When starting from the initial condition such that $x=0$\,, $E=0.1$ and $-5<y<5$\,, some of motions escape $B$ at $t=70$\,. 
Figure \ref{matrix stable} shows the remaining initial conditions in the $(y,p_y)$-plane.

\medskip

\begin{figure}[htpb]
\begin{minipage}[ht]{0.55\linewidth}
\centering
\includegraphics[width=0.9\linewidth]{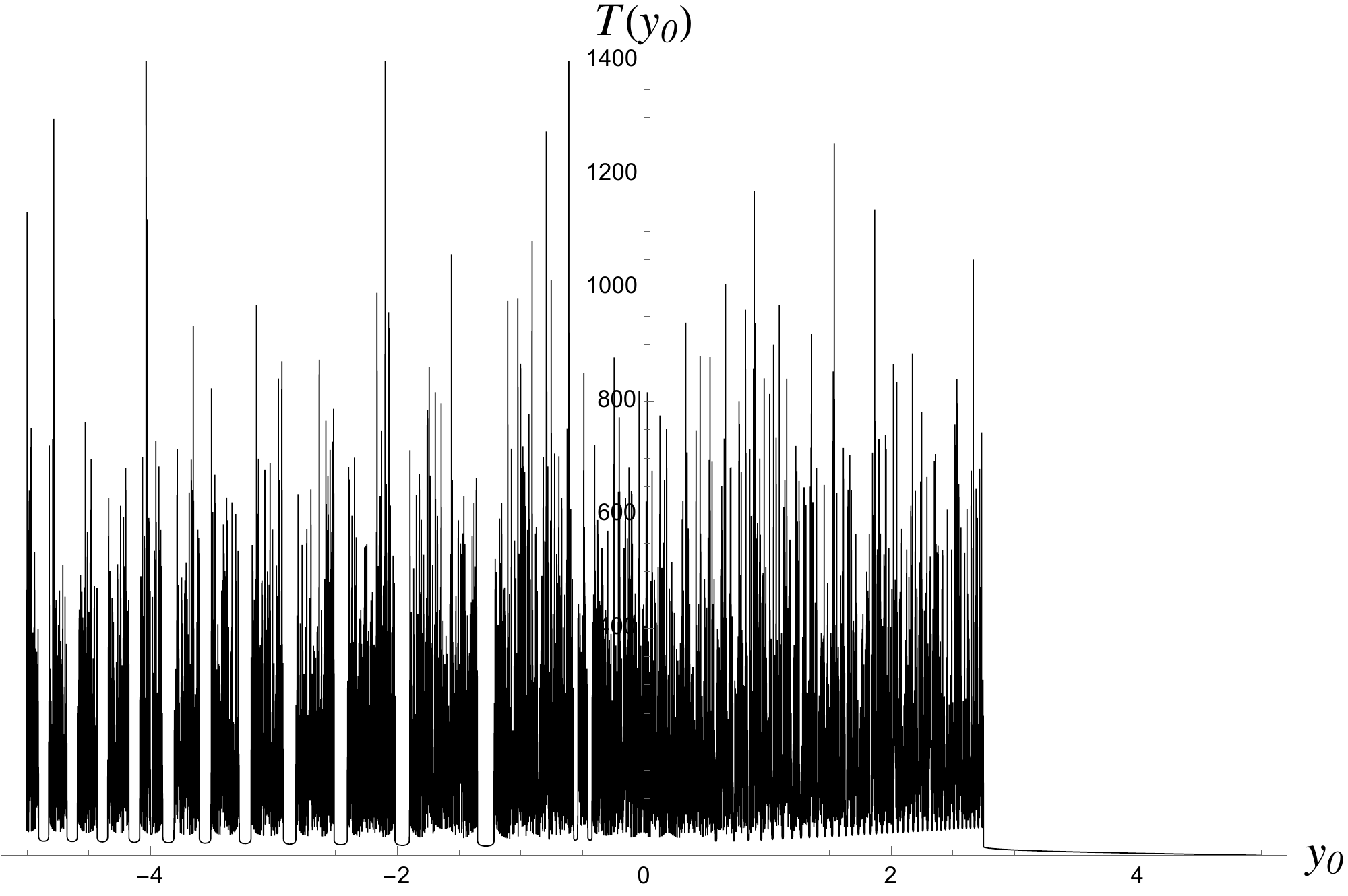}
\vspace*{0.5cm} 

\includegraphics[width=0.9\linewidth]{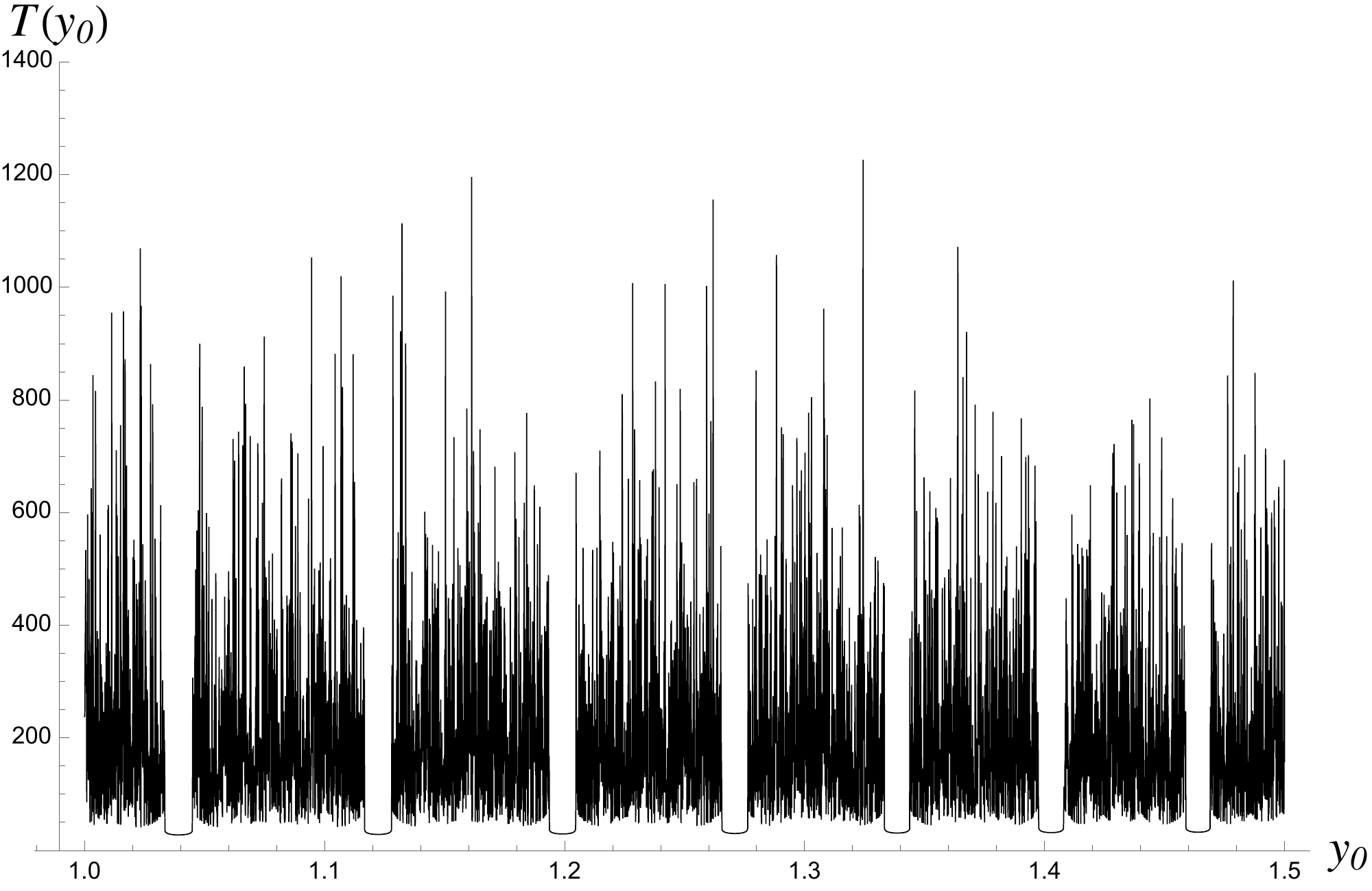}

\vspace*{0.5cm}
\includegraphics[width=0.9\linewidth]{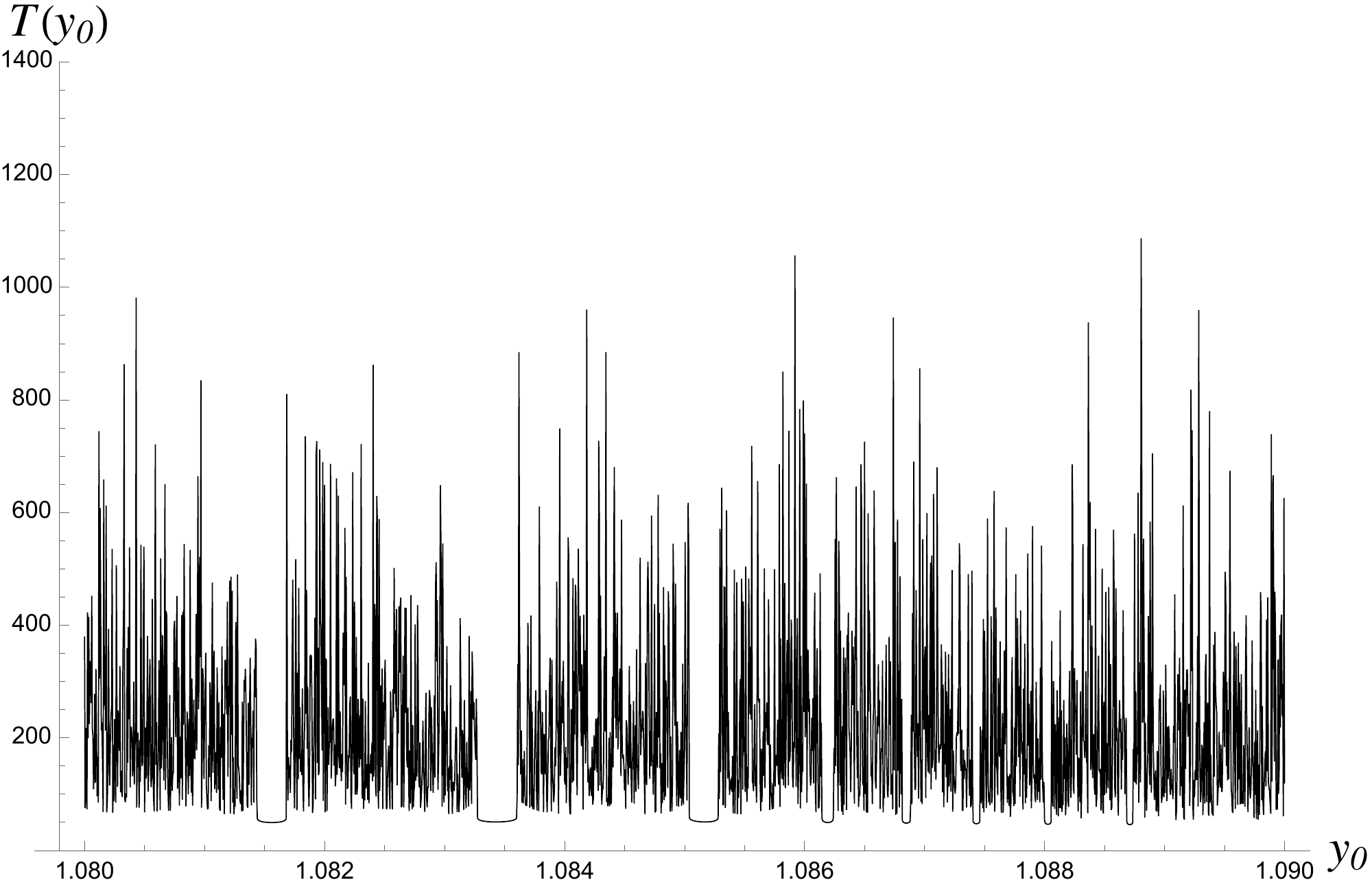}
\caption{\footnotesize Time delay function for $x=0$\,, $p_{y}=0.3$\,, $E=0.1$\,.
Here $y_0$ is an initial value of $y$\,.
The second is a zoom-up of the range $1.0<y_0<1.5$\,.
The third is a zoom-up of the range $1.08<y_0<1.09$\,.}\label{fig:matrix time}
\end{minipage}
\begin{minipage}[ht]{0.45\linewidth}
\centering
\includegraphics[width=1\linewidth]{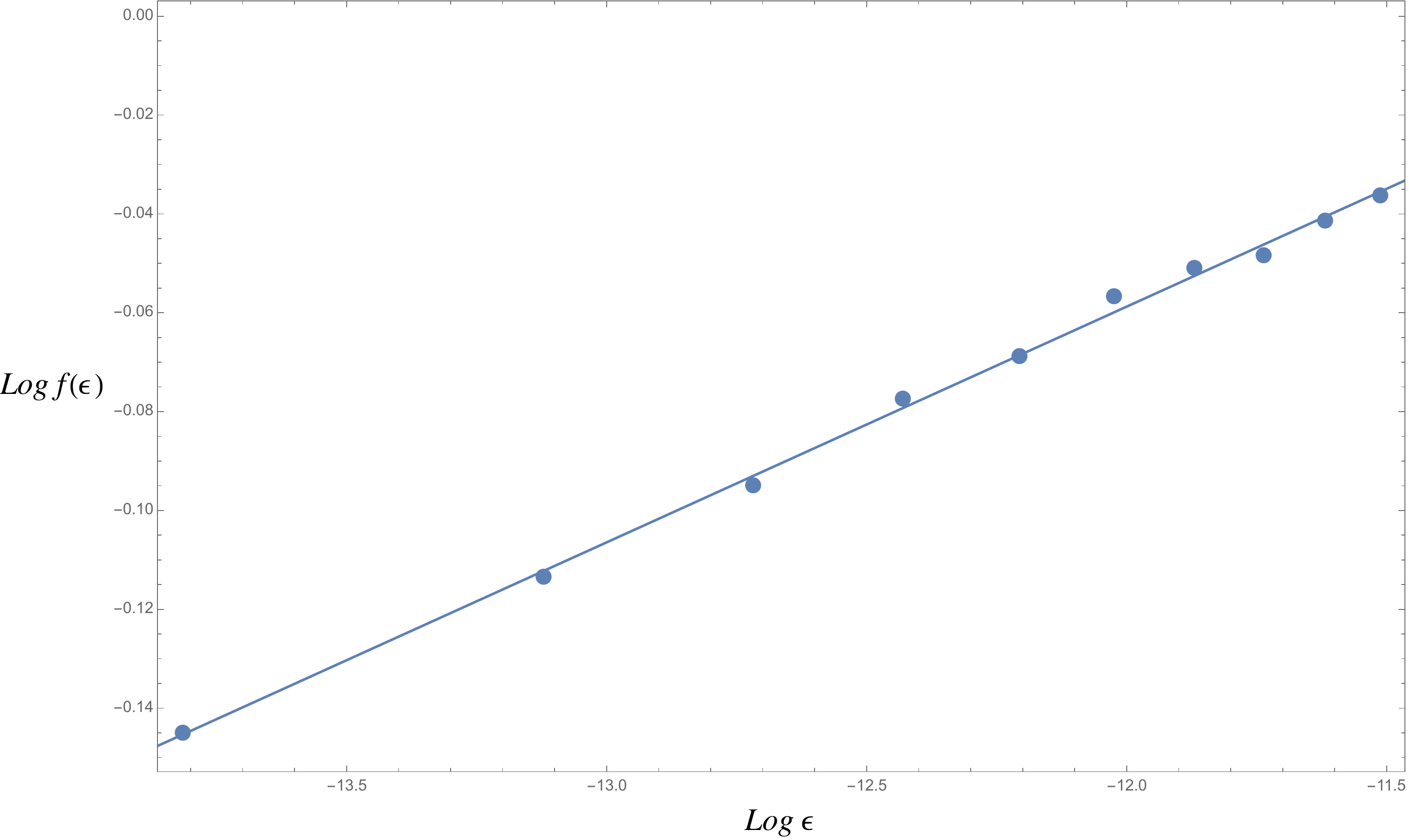} 
\vspace{0.5cm}
\caption{\footnotesize The uncertain fraction with $h=1$\,.
The axes are given by the natural logarithms.}\label{fig:matrix dimension}
\end{minipage}
\end{figure}

Then, we have computed the time delay function for the initial condition $x=0$ and $p_y=0.3$ (as shown in Fig.\,\ref{fig:matrix time}). 
There are singularities and smooth intervals as in the four-hill potential case. By zooming up to a part of the graph, a self-similar structure can be observed. The singularities form a Cantor-like set again and its fractal dimension can be estimated by using the uncertain fraction.  

\medskip 

The uncertain dimension is computed based on Fig.\,\ref{fig:matrix dimension},
where we took $10^4$ reference points $x_0$\,.
As a result, the uncertain fraction is fitted as
\begin{align}
\log f(\epsilon)=(0.0465\pm 0.0007)\log\epsilon+(0.500\pm 0.009)\,,
\end{align}
and thus the fractal dimension is given by
\begin{align}
d=0.9535\pm 0.0007\,.
\end{align}
After all, the singularities of the time delay function under the ansatz (\ref{sphere ansatz}) indeed exhibit a fractal structure. This indicates that the membrane instability is identified with chaotic instability.  

\medskip

For the bosonic matrix model, quantum corrections lift up all of the flat directions and bosonic membranes are stable. But in the case of supermembranes, the quantum corrections vanishes due to the presence of supersymmetries. Hence the supermembrane is unstable 
in the same manner as in the case of bosonic membranes at classical level. Thus, the chaotic instability discussed above is common to the supermembrane at quantum level.


\subsection*{Cutoff-dependence}

\begin{figure}[htpb]
\centering
\includegraphics[width=0.7\linewidth]{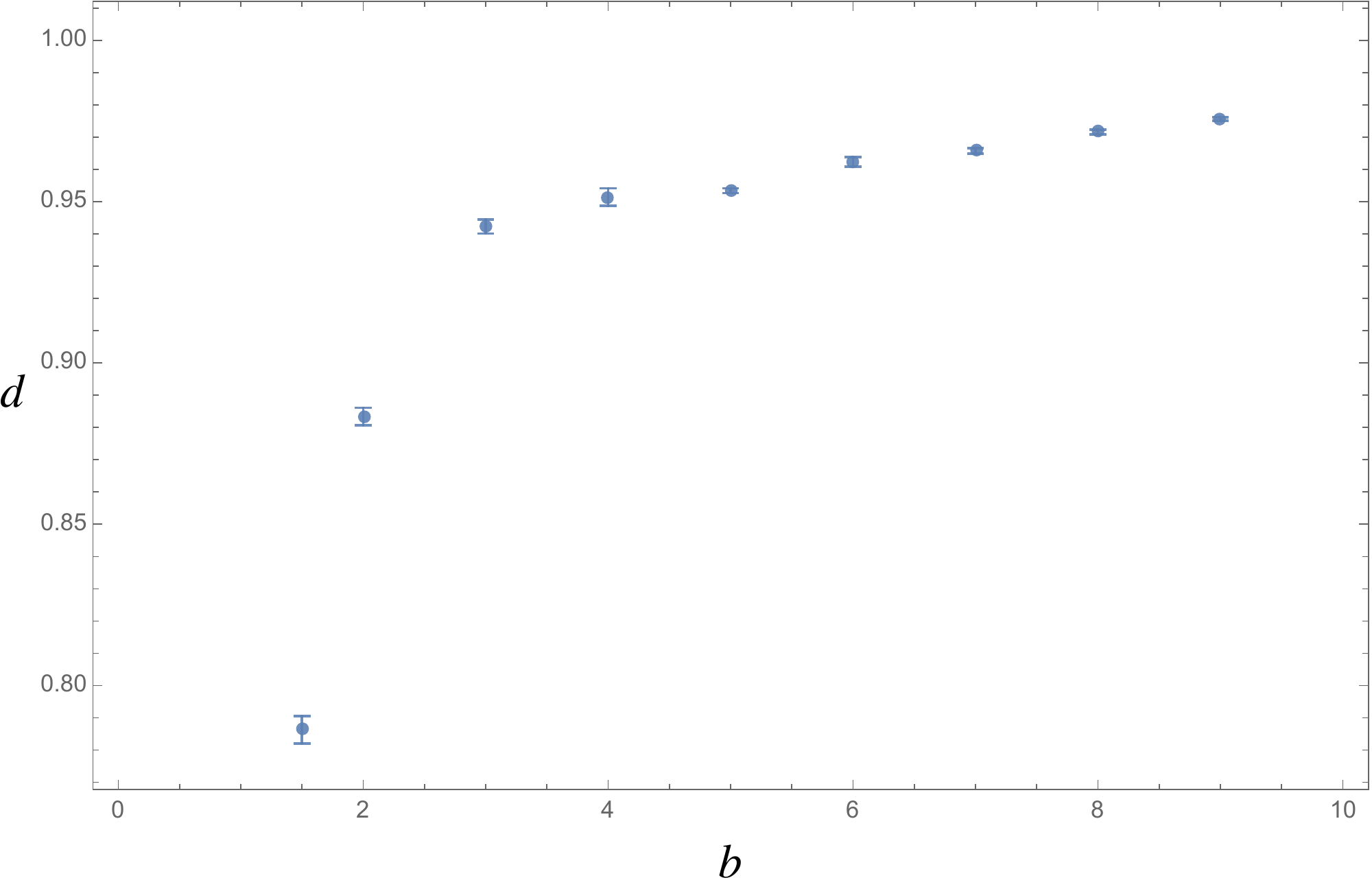}
\caption{\footnotesize The $B$-dependence of the uncertain dimension $d$\,.
The size of the chaotic region $B$ is defined by $|x|\leq b$ and $|y|\leq b$\,.
}\label{fig:cutoff_dependence}
\end{figure}

So far, we have discussed the fractal structure and computed its fractal dimension by fixing the size of the chaotic region $B$ as $|x|<5$ and $|y|<5$\,.
Since the motion is not chaotic outside the chaotic region $B$, it is expected that the fractal structure would not change if $B$ is sufficiently large. The $B$-dependence is plotted in Fig.\,\ref{fig:cutoff_dependence}, where we have  calculated the fractal dimension varying $b$ such that $B$ is given by $|x|<b$ and $|y|<b$\,, while the other conditions are fixed. It seems likely that $d$ is saturated around $d\simeq 0.96$ and appears to asymptote to a universal value in the limit of $b\to\infty$\,. However, this limit is ideally possible because plateaus in the time delay function become smaller as $b$ increases and thus the fitting for the uncertain fraction $f(\epsilon)$ does not work well. Therefore, it is very difficult to actually conclude whether the value of $d$ is really saturated or gradually increasing towards 1.

\section{Conclusion and Discussion}\label{sec:conclusion}

We have considered chaotic scattering in the BFSS matrix model. The membrane instability associated with the flat directions in the potential is identified as chaotic scattering. For a reduced system, we have shown a fractal structure in the space of initial conditions for which a certain type of membrane moves in the chaotic region. In addition, we have shown that the singularities in the time delay function form a Cantor-like set and estimated its fractal dimension. Thus a fractal structure in the membrane decay process has been revealed.  

\medskip 

Our analysis here has been carried out with fixed coupling constant and a certain value of the energy. The strength of transient chaos varies depending on values of the coupling constant and energy. Elucidating the dependence of the fractal dimension on the coupling constant and energy is an interesting problem (For a similar direction, see a recent interesting work \cite{Hashimoto:2021afd}). For this issue, we will report in another place in the near future. 

%

\medskip 

There are other questions. We have discussed here chaotic scattering at classical level. It is interesting to take quantum corrections into account. In the bosonic matrix model, all of the flat directions are lifted up by quantum corrections. However, in the supermembrane case, the quantum corrections are cancelled out by supersymmetries and the instability survives. The ground-state wave function of a supermembrane is constructed in \cite{Piljin}. It is interesting to study the relation between this wave function and the fractal structure revealed in this paper. In general, the self-similar structure in fractal is truncated at the Planck scale at quantum level. It is significant to figure out this truncation mechanism.  

\medskip 

It is also interesting to look for applications of the fractal structure in the context of the gauge/gravity correspondence. It may be related to graviton scattering in eleven-dimensional supergravity. The results obtained here is also common to the four-dimensional classical Yang-Mills theory. Therefore, the corresponding phenomenon should exist on AdS$_5$\,, though we have no idea what it is indeed. We hope that our fractal result will shed light on a new aspect of the gauge/gravity correspondence.

\subsection*{Acknowledgments}

We are very grateful to Koji Hashimoto for useful comments and discussions.
The work of O.\,F.\ was supported by Grant-in-Aid for JSPS Fellows No.~21J22806.
The work of K.Y.\ was supported by 
JSPS Grant-in-Aid for Scientific Research (B) No.\,18H01214.

%


\end{document}